\documentclass[twocolumn]{28onr_art}
\usepackage{natbib}
\usepackage{ulem}
\usepackage{graphicx}
\usepackage{fixltx2e}
\usepackage{color}
\usepackage[colorlinks=true,
            urlcolor=blue,
           linkcolor=docgreen,
            citecolor=docgreen,
            bookmarks=true,
            pdftitle={Numerical Prediction of a Seaway},
            pdfauthor={Douglas G. Dommermuth},
            pdfsubject={Proceedings of the 28th Naval Hydrodynamics Symposium},
            pdfkeywords={Breaking Waves, Nonlinear Waves, HOS, NFA, FSM},
	   bookmarksopen=false,
            pdfpagemode=UseNone]{hyperref}

\definecolor{docgreen}{rgb}{0,.5,0}

\pdfoutput = 1

\def\curvered{\mbox{(~\textcolor{red}{---------------}~)}}
\def\curvegreen{\mbox{(~\textcolor{green}{---------------}~)}}
\def\curveblue{\mbox{(~\textcolor{blue}{---------------}~)}}

\def\curve{\mbox{(~---------------~)}}

\def\curuud{\mbox{(~--~--~--~--~--~--~)}}

\def\be{\begin{eqnarray}}
\def\ee{\end{eqnarray}}
\def\benl{\begin{eqnarray*}}
\def\eenl{\end{eqnarray*}}

\newcommand{\nwc}{\newcommand}
\nwc{\bm}{\boldmath}
\nwc{\m}{\mbox}
\nwc{\ubm}{\unboldmath}
\nwc{\bmU}{\m{\bm$U$\ubm}}
\nwc{\bmX}{\m{\bm$X$\ubm}}
\nwc{\bmu}{\m{\bm$u$\ubm}}
\nwc{\bmx}{\m{\bm$x$\ubm}}
\nwc{\bmz}{\m{\bm$z$\ubm}}
\nwc{\bmv}{\m{\bm$v$\ubm}}
\nwc{\bmw}{\m{\bm$w$\ubm}}
\nwc{\bmW}{\m{\bm$W$\ubm}}
\nwc{\bmn}{\m{\bm$n$\ubm}}
\nwc{\bmG}{\m{\bm$G$\ubm}}
\nwc{\bmF}{\m{\bm$F$\ubm}}
\nwc{\bmI}{\m{\bm$I$\ubm}}
\nwc{\bmN}{\m{\bm$N$\ubm}}
\nwc{\bmP}{\m{\bm$P$\ubm}}
\nwc{\bmcalP}{\m{\bm $\cal P$\ubm}}
\nwc{\bmV}{\m{\bm$V$\ubm}}
\nwc{\bmS}{\m{\bm$S$\ubm}}

\begin{document}

\title{Numerical Prediction of a Seaway}

\author{Douglas G.\ Dommermuth$^1$, Thomas C.\  Fu$^2$, Kyle A.\ Brucker$^1$, Thomas T.\ O'Shea$^1$, and Donald C.\ Wyatt$^1$}

\affiliation{\small $^1$Naval Hydrodynamics Division, Science Applications International Corporation,
\\ 10260 Campus Point Drive, MS C5, San Diego, CA  92121, USA \\
$^2$Naval Surface Warfare Center, Carderock Division, \\ 9500 MacArthur Blvd, West Bethesda, MD 20817, USA}

\maketitle

\begin{abstract}
The ability of three wave theories to predict statistics and the crest kinematics of a seaway is quantified.  The three wave theories are high-order spectral (HOS) theory, free-surface mapping (FSM), and volume-of-fluid (VOF).  Issues associated with applying these methods are discussed, including free-surface adjustment, smoothing and filtering, and forcing.    Two long-crested regular waves with varying bandwidth and moderate steepness are used to benchmark the performance of the wave theories.   As a more stringent test, a broad-banded long-crested seaway is simulated.
\end{abstract}

\section{Introduction}

The need for US Navy surface combatant vessels to safely operate in extreme sea states requires the characterization and simulation of this environment and the development of suitable models of the wave field. To date, the current open-ocean wave-field models such as WAM are focused on the air-sea energy exchange, not the kinematics that would drive the forcing a ship would experience in those sea states \citep{wam88}. Additionally, current CFD simulations of the wave field are too computationally expensive to be coupled with design tools, so there exists a need for computationally fast and accurate wave kinematic models.

The ability of three wave theories to predict the nonlinear statistics and the crest kinematics of a seaway is quantified.  The three wave theories are high-order spectral theory (HOS), free-surface mapping (FSM), and volume-of-fluid (VOF) interface capturing.   The Numerical Flow Analysis (NFA) code provides a framework for testing VOF's ability to model nonlinear wave interactions.   Details of the HOS, FSM, and NFA formulations are provided respectively in \citet{dommermuth87}, \citet{dommermuth94}, and \citet{nfa1}.  HOS and FSM are single-valued free-surface approximations, whereas NFA allows wave overturning and wave breaking.   HOS uses perturbation approximations and Taylor series approximations to solve the nonlinear free-surface boundary conditions.    FSM maps the free surface to a flat plane where the fully nonlinear free-surface boundary conditions are solved directly.  NFA is a two-phase model that simulates the flow in the air and water simultaneously with no approximation.  HOS is a surface discretization.   FSM, as formulated in this paper, is a volume discretization.   NFA solves the full three-dimensional Navier-Stokes equations using an implicit sub-grid scale (SGS) model.  FSM and NFA provide the wave kinematics throughout the water column.   HOS requires additional computations to predict wave kinematics beneath the free surface.   HOS has spectral accuracy in space and 4th-order accuracy in time.    FSM is fourth-order accurate in space and time.   NFA is second-order accurate in space and time.   HOS is the most efficient computationally.   FSM is less efficient than HOS, but more efficient than NFA.   For simulating seaways, HOS and FSM are limited due to their single-valued free-surface approximation.  FSM has a greater range of validity than HOS because FSM makes fewer approximations.   The range of validity of NFA is not restricted, but NFA is computationally intensive.

The ability of HOS, FSM, and NFA to simulate free-surface waves is quantified using regular and irregular waves.   Let $H$ and $\lambda$ denote respectively wave height and wave length.  Two long-crested regular waves with moderate steepness ($H_1/\lambda_1$=$H_2/\lambda_2$=0.025) and with bandwidths ($\lambda_1/\lambda_2$) varying from 2 to 32 are used to benchmark the basic performance of the wave theories.  As a more stringent test, simulations of a short-crested seaway are performed using a Pierson-Moskowitz spectrum as a basis.  The nonlinear theories are initialized using adjustment procedures \citep{dommermuth00}.  For very high sea states, filtering in physical and wavenumber space is used to extract wave energy in regions where the waves get too steep to model using HOS and FSM.    NFA requires filtering to prevent tearing of the free-surface interface where there is jump in tangential velocity between water and air.   The total wave energy is conserved in HOS, FSM, and NFA simulations using forcing that is uniform over all wavenumbers and equally partitioned between kinetic and potential energies.   For HOS and FSM, the numerical simulations are run until the relevant wave statistics are stationary.   The ability of NFA to model subtle wave-wave interactions is illustrated.   For HOS and FSM, statistics are calculated for the free-surface elevation and the surface water-particle velocity.   Depending on sea state, the advantages of disadvantages of each theory in terms of providing input to ship-motions codes like TEMPEST/FREDYNE are discussed.

\section{\label{sec:formulation} Formulation}

\subsection{\label{sec:bvp}Boundary-value problems for spectral and mapping methods}

Laplace's equation is satisfied within the fluid:
\begin{eqnarray}
\label{laplace}
\phi_{xx}+\phi_{yy}+\phi_{zz}=0 \;\; {\rm for} \;\; -h \leq z \leq \eta \;\; ,
\end{eqnarray}
\noindent where $\phi({\bf x},z,t)$ is the velocity potential, $\eta({\bf x},t)$ is the free-surface elevation, and $h$ is the bottom. ${\bf x}=(x,y)$ is a vector in the horizontal plane, and $t$ denotes time.  We assume that $\eta$ is continuous and single valued.

Following \citet{Zakaharov}, $\phi^s$ is the potential evaluated on the free surface:
\begin{eqnarray}
\phi^s({\bf x},t)=\left. \phi({\bf x},z,t) \right|_{z=\eta} \;\; .
\end{eqnarray}
The temporal and spatial derivatives of $\phi^s$ in terms of $\phi$ are 
\begin{eqnarray}
\frac{\partial \phi^s}{\partial t} & = & \left. \frac{\partial \phi}{\partial t} \right|_{z=\eta}+ \frac{\partial \eta}{\partial t} \left. \frac{\partial \phi}{\partial z}  \right|_{z=\eta}  \nonumber \\
\nabla_x \phi^s & = & \left. \nabla_x \phi \right|_{z=\eta} + \nabla_x \eta \; \left. \frac{\partial \phi}{\partial z} \right|_{z=\eta}  \;\; ,
\end{eqnarray}
where $\nabla_x$ denotes the horizontal gradient:
\begin{eqnarray}
\nabla_x \equiv \left( \frac{\partial}{\partial x}, \frac{\partial}{\partial y} \right) \;\; .
\end{eqnarray}

Length and velocity scales are respectively normalized by $L_o$ and $U_o$.   Based on this normalization, the dynamic and kinematic free-surface boundary conditions with weak viscous effects are
\begin{eqnarray}
\label{fsbc}
\frac{\partial \phi^s}{\partial t} & = & - \frac{1}{F_r^2} \eta -\frac{1}{2} \nabla_x \phi^s \cdot \nabla_x \phi^s \nonumber \\
& + & \frac{1}{2} \left(1+\nabla_x \eta \cdot \nabla_x \eta \right) \frac{\partial \phi}{\partial z}^2 \nonumber \\
& + & \frac{2}{Re} (\frac{\partial^2 \phi}{\partial x^2} + \frac{\partial^2 \phi}{\partial y^2} ) \nonumber \\
& - & \frac{1}{W_e} \nabla \cdot {\bf n} \;\; {\rm on \; z=\eta} \nonumber \\
\frac{\partial \eta}{\partial t} & = & -\nabla_x \phi^s \cdot \nabla_x \eta \nonumber \\
& + &  ( 1+ \nabla_x \eta \cdot \nabla_x \eta )  \frac{\partial \phi}{\partial z} \nonumber \\ 
& + & \frac{2}{Re} (\frac{\partial^2 \eta}{\partial x^2} + \frac{\partial^2 \eta}{\partial y^2} ) \;\; {\rm on \; z=\eta} \;\; ,
\end{eqnarray}
\noindent where $F_r=U_o/\sqrt{g L_o}$, $R_e=U_o L_o / \nu$, and $W_e=\rho U_o^2 L_o/\sigma$ are respectively the Froude, Reynolds, and Weber numbers.  $g$ is the acceleration of gravity, 
$\nu$ is the kinematic viscosity, $\rho$ is the density of water, and $\sigma$ is the surface tension.  ${\bf n}$ is the unit normal that points out of the fluid. 

\subsection{\label{sec:spectral}Formulation for spectral method}

Following \citet{dommermuth87}, we assume that $\phi$ and $\eta$ are O($\epsilon$) quantities, where $\epsilon$, a small parameter, is the wave steepness.   The potential is expanded in a perturbation series up to order $M$ in $\epsilon$:
\begin{eqnarray}
\phi({\bf x},z,t) = \sum_{m=1}^{M} \phi^{(m)}({\bf x},z,t)  \;\; .
\end{eqnarray}
Each perturbation potential is further expanded in a Taylor series about $z=0$ to evaluate the surface potential:
\begin{eqnarray}
\label{perturb}
\lefteqn{\phi^s({\bf x},t) = \phi({\bf x},\eta,t)} & & \nonumber \\
 & & = \sum_{m=1}^{M} \sum_{k=0}^{M-m} \frac{\eta^k} {k!} \frac{\partial^k}{\partial z^k} \phi^{(m)}({\bf x},0,t) \;\; .
\end{eqnarray}
At any instant of time, $\phi^s$ and $\eta$ are known, so that (\ref{perturb}) provides a Dirichlet condition for the unknown $\phi^{(m)}$.   By collecting terms at each order, a sequence of boundary-value problems follows:
\begin{eqnarray}
\phi^{(m)}({\bf x},0,t) & = & R^{(m)} \; , \;\;  {\rm m=1,2,3, \ldots, M} \nonumber \\
R^{(1)} & = & \phi^s({\bf x},t) \nonumber \\
R^{(m)} & = & -\sum_{k=1}^{m-1} \frac{\eta^k} {k!} \frac{\partial^k}{\partial z^k} \phi^{(m-k)}({\bf x},0,t) \; , \nonumber \\
& & {\rm m=2,3, \ldots, M}
\end{eqnarray}
To solve the boundary-value problems, each $\phi^{(m)}$ is expanded in terms of a finite number ($N$) of  eigenfunctions ($\Psi_n$):
\begin{eqnarray}
\phi^{(m)}({\bf x},z,t) = \sum_{n=1} ^N \phi_n^{(m)}(t) \Psi_n({\bf x},z) \; , \; z \leq 0 \;\; .
\end{eqnarray}-
The vertical velocity evaluated on $z=\eta$ is
\begin{eqnarray}
\label{zvelo}
\lefteqn{\phi_z({\bf x},\eta,t)} & & \nonumber \\
 & & = \sum_{m=1}^{M} \sum_{k=0}^{M-m} \frac{\eta^k} {k!}  \sum_{n=1} ^N \phi_n^{(m)}(t)  \frac{\partial^{k+1}}{\partial z^{k+1}} \Psi({\bf x},0)  \;\; . \nonumber \\
 & & 
\end{eqnarray}
For constant finite depth, 
\begin{eqnarray}
\label{eigen}
\Psi_n ({\bf x},t) = \frac{\cosh\left[ | {\bf k}_n| (z+h) \right]}{\cosh( |{\bf k}_n| h )} \exp(\imath {\bf k}_n \cdot {\bf x}) \;\; ,
\end{eqnarray}
\noindent where ${\bf k}_n=(k_x, k_y)$.    Equations \ref{eigen} and \ref{zvelo} are substituted into the free-surface boundary conditions \ref{fsbc}, and the resulting equations are solved using a pseudo-spectral method.   Details are provided in \citet{dommermuth87}.   Other examples of HOS formulations with applications are provided by \citet{wu04} and \citet{blondel08}.

\subsection{\label{sec:mapped}Formulation for free-surface mapping}

Instead of using a combination of perturbation and Taylor-series expansions like the spectral formulation, the boundary-value problem is solved directly using a mapping.    For $-h \leq z \leq \eta(x,y)$, the free surface is mapped to a flat plane:
\begin{eqnarray}
\zeta = \frac{z+h}{h+\eta(x,y)} \;\; ,
\end{eqnarray}
\noindent where $\zeta=1$ for $z=\eta$, and $\zeta=0$ for $z=-h$.  Let $\Phi$ denote the velocity potential in the mapped coordinate system:
\begin{eqnarray}
\phi(x,y,z)= \Phi(x,y,\zeta) \;\; .
\end{eqnarray}
The first derivatives of $\phi$ in terms of $\Phi$ are
\begin{eqnarray}
\label{deri1}
\phi_x & = & \Phi_x +\zeta_x \Phi_\zeta \nonumber \\
\phi_y & = & \Phi_y +\zeta_y \Phi_\zeta \nonumber \\
\phi_z & = & \zeta_z \Phi_\zeta \;\; .
\end{eqnarray}
Similarly, the second derivatives are
\begin{eqnarray}
\label{deri2}
\phi_{xx} & = & \Phi_{xx}+2 \zeta_x \Phi_{x\zeta} +\zeta_{xx} \Phi_\zeta+\zeta^2_x \Phi_{\zeta \zeta} \nonumber \\
\phi_{yy} & = & \Phi_{yy}+2 \zeta_y \Phi_{y\zeta} +\zeta_{yy} \Phi_\zeta+\zeta^2_y  \Phi_{\zeta \zeta} \nonumber \\
\phi_{zz} & = &  \zeta_z^2 \Phi_{\zeta\zeta} \;\; .
\end{eqnarray}
Substituting \ref{deri2} into \ref{laplace} gives Laplace's equation in a mapped coordinate system:
\begin{eqnarray}
\label{mapped_bvp}
  \lefteqn{\Phi_{xx}+\Phi_{yy} +\zeta_z^2 \Phi_{\zeta\zeta}+2 \zeta_x \Phi_{x\zeta} +2 \zeta_y \Phi_{y\zeta}} & & \nonumber \\
  & & +(\zeta_{xx} +\zeta_{yy} ) \Phi_\zeta+(\zeta^2_x+\zeta^2_y )\Phi_{\zeta\zeta}=0 \;\; .
\end{eqnarray}
Equation \ref{mapped_bvp} is solved using a preconditioned 4th-order finite-difference scheme.   The equation for $\phi_z$ in \ref{deri1} is substituted into \ref{fsbc} and the free-surface boundary conditions are evolved in time.  Details of a similar two-dimensional formulation are provided in \citet{dommermuth94}.   \citet{clamond01} use Green's theorem with spectral basis functions to solve the fully-nonlinear free-surface boundary conditions without using a volume discretization.

\subsection{\label{sec:vof}Formulation for Volume of Fluid (VOF) method}

Consider the  immiscible turbulent flow at the interface between air and water with $\rho_a$ and $\rho_w$ respectively denoting the densities of air and water.   Similar to the potential-flow approaches, physical quantities are normalized by characteristic velocity ($U_o$), length ($L_o$),  time ($L_o/U_o$), density ($\rho_w$), and pressure ($\rho_w U_o^2$) scales.  

Let $\alpha$ denote the fraction of fluid that is inside a cell. By definition, $\alpha=0$ for a cell that is totally filled with air, and $\alpha=1$ for a cell that is totally filled with water.   In terms of $\alpha$, the normalized density is express as
\begin{eqnarray}
\label{alpha}
\rho(\alpha) & = & \lambda + (1 - \lambda) \alpha \;\; ,
\end{eqnarray}
\noindent where $\lambda = \rho_a/\rho_w$ is the density ratio between air and water.  

Let $u_i$ denote the normalized three-dimensional velocity field as a function of normalized space ($x_i$) and normalized time ($t$).  The conservation of mass is
\begin{eqnarray}
\label{mass}
\frac{\partial \rho}{\partial t} +\frac{\partial u_j \rho}{\partial x_j} = 0 \;\; .
\end{eqnarray}

For incompressible flow,
\begin{eqnarray}
\label{density}
\frac{\partial \rho}{\partial t} +u_j \frac{\partial  \rho}{\partial x_j} = 0 \;\; .
\end{eqnarray}

Subtracting Equation (\ref{density}) from (\ref{mass}) gives a solenoidal condition for the velocity:
\begin{eqnarray}
\label{solenoidal}
\frac{\partial u_i}{\partial x_i} = 0 \;\; .
\end{eqnarray}

Substituting Equation (\ref{alpha}) into (\ref{mass}) and making use of (\ref{solenoidal}), provides an advection equation for  the volume fraction:
\begin{eqnarray}
\label{vof}
\frac{\partial \alpha}{\partial t}+ \frac{\partial}{\partial x_j} \left(u_j \alpha \right)= 0 \;\; .
\end{eqnarray}

For an infinite Reynolds number, viscous stresses are negligible, and the conservation of momentum is
\begin{eqnarray}
\label{momentum}
\frac{\partial u_i}{\partial t}+\frac{\partial}{\partial x_j} \left(u_j u_i \right)  =  -\frac{1}{\rho} \frac{\partial p}{\partial x_i} -\frac{p_s}{\rho} \frac{\partial H(\alpha)}{\partial x_i} - \frac{\delta_{i3}}{F_r^2}  \;\; ,
\end{eqnarray}
\noindent where $F_r^2 = U_o^2/(g L_o)$ is the Froude number, and $g$ is the acceleration of gravity.  $p$ is the pressure and $p_s$ is a stress that acts normal to the interface.  $H(\alpha)$ is a Heaviside function, and $\delta_{ij}$ is the Kronecker delta function.   

The divergence of the momentum equations (\ref{momentum}) in combination with the solenoidal condition (\ref{solenoidal}) provides a Poisson equation for the dynamic pressure:
\begin{eqnarray}
\label{poisson} 
\frac{\partial}{\partial x_i} \frac{1}{\rho} \frac{\partial
p}{\partial x_i} = \Sigma \;\; ,
\end{eqnarray}
\noindent where $\Sigma$ is a source term.  The pressure is used to project the velocity onto a solenoidal field.  Details of the volume fraction advection, the pressure projection, and the numerical time integration are provided in \citet{dommermuth07} and \citet{dommermuth08}. Sub-grid scale stresses are modeled using an implicit model that is built into the treatment of convective terms.  The performance of the implicit SGS model is provided in \citet{nfa3}. 

\subsection{Smoothing}

Smoothing is required in HOS and FSM simulations of broad-banded wave spectra.   Smoothing prevents the pileup of energy at high wave numbers.     Filtering is required in NFA simulations to prevent tearing of the free surface due to a discontinuity in the tangential velocity between air and water across the free surface.   The details of the smoothing and filtering algorithms are provided in the next three sections.

\subsubsection{\label{sec:smooth_spectral}Smoothing for spectral method}

Filtering in wavenumber space is used for the spectral method.
\begin{eqnarray}
\label{smooth_spectral}
F_{m,n} (\gamma_c) =
\left\{
\begin{array}{lr}
1 & (\frac{k_m}{\gamma_c k_M})^2+(\frac{k_n}{\gamma_c k_N})^2 \le 1\\
   & \\
0 & (\frac{k_m}{\gamma_c k_M})^2+(\frac{k_n}{\gamma_c k_N})^2 > 1\\
\end{array} \;\; ,
\right.
\end{eqnarray}
\noindent where $k_m$ and $k_n$ are respectively the wavenumbers along the $x$ and $y-$axes, $k_M$ and $k_N$ are the corresponding Nyquist wavenumbers, and $0 < \gamma_c \leq 1$ is the cut-off parameter.  Equation \ref{smooth_spectral} is applied to $\phi^s$ and $\eta$ every time step.

\subsubsection{\label{sec:smooth_mapping}Smoothing for mapping method}

An eleven-point smoothing scheme in physical space is used for the free-surface mapping formulation.  The smoothing is applied sequentially along the $x$ and $y-$axes.   The stencil is
\begin{eqnarray}
\label{smooth_map}
\lefteqn{\tilde{F}_i =\frac{193}{256} F_i 
+ \frac{105}{512} \left( F_{i-1} + F_{i+1} \right)}  & & \nonumber \\
& - &  \frac{15}{128} \left( F_{i-2} + F_{i+2} \right) 
+ \frac{45}{1024} \left( F_{i-3} + F_{i+3} \right) \nonumber \\ 
& - & \frac{5}{512} \left( F_{i-4} + F_{i+4} \right) 
+ \frac{1}{1024} \left( F_{i-5} + F_{i+5} \right) \;\; , \nonumber \\
\end{eqnarray}
where $F_i$ is a discrete function at index $i$, and $\tilde{F}_i$ is the smoothed discrete function.  Equation \ref{smooth_map} is applied to $\phi^s$ and $\eta$ every 5 time steps.

\subsubsection{\label{sec:smooth_vof}Smoothing for VOF method}

The free-surface boundary layer is not resolved in VOF simulations at high Reynolds numbers with large density jumps such as air and water.  Under these circumstances, the tangential velocity is discontinuous across the free-surface interface and the normal component is continuous.   As a result, unphysical tearing of the free surface tends to occur.   Favre-like filtering can be used to alleviate this problem by forcing the air velocity at the interface to be driven by the water velocity in a physical manner.    Consider the following projection,
\begin{eqnarray}
\label{smooth_velo}
\tilde{u}_i=\frac{\left< \rho u_i  \right>}{\left< \rho \right>} \;\; {\rm for} \; \alpha \ge 0.5 \;\; ,
\end{eqnarray}
where $\tilde{u}_i$ is the smoothed velocity field, $u_i$ is the unfiltered velocity field, $\rho$ is the density, and $\alpha$ is the volume fraction.  Brackets denote smoothing.
\begin{eqnarray}
\left< F({\bf x}) \right> = \int_{\rm v_\xi} W({\bf x_\xi}) F({\bf x}-{\bf x_\xi}) {\rm d v_\xi}  \;\; .
\end{eqnarray}
Here, $F(x)$ is a general function, ${\rm v}$ is a control volume that surrounds a cell, and $W(x)$ is a weighting function that neither overshoots or undershoots the maximum or minimum allowable density.  Due to the  high density ratio between water and air, Equation~\ref{smooth_velo} tends to push the water-particle velocity into the air.   Once the velocity is filtered, we need to project it back onto a solenoidal field in the fluid volume (V).
\begin{eqnarray}
\label{project1}
u_i = \tilde{u}_i - \frac{1}{\rho} \frac{\partial \psi}{\partial x_i}  \;\; {\rm in \; V} \; ,
\end{eqnarray}
where $\psi$ is a potential function.     For an incompressible flow, we require that $u_i$ is solenoidal care of Equation~({\ref{solenoidal}).  Substituting (\ref{project1}) into (\ref{solenoidal}) gives a Poisson equation for $\psi$:
\begin{eqnarray}
\label{project2}
\frac{\partial}{\partial x_i} \frac{1}{\rho} \frac{\partial \psi}{\partial x_i} = \frac{\partial \tilde{u}_i}{\partial x_i}   \;\; {\rm in \; V} \; .
\end{eqnarray}
\noindent We typically apply the filtering every 20 time steps.  Details of the implementation of the preceding filter are provided in \citet{nfa6}.

\subsection{Energy Pumping}

As a result of smoothing, filtering, and actual wave breaking, energy is not conserved in numerical simulations of free-surface waves.    Aside from the need to compensate for these effects, numerical simulations also require energy input to model the effects of wind acting on the ocean surface.     The methods that are describe in the next three sections provide a mechanism for pumping energy into numerical simulations without adversely affecting the shape of the wave spectrum.

\subsubsection{Pumping for spectral and mapped methods}

The total energy as a function of time ($E(t)$) is
\begin{eqnarray}
\lefteqn{E(t)=\int_{S_o} ds \, \phi^s \eta_t + \frac{1}{F_r^2} \int_{S_o} ds \, \eta^2} & &  \nonumber \\ 
& + & \frac{1}{W_e} \int_{S_o} ds \, \left( \sqrt{1+\eta_x^2 + \eta_y^2} -1 \right) \;\; ,
\end{eqnarray}
\noindent where the first term on the right-hand side is the kinetic energy, the second term is the potential energy, and the last term is the superficial energy.  $S_o$ is the horizontal plane.  Due to the effects of smoothing, the total energy will decrease over time.  The total energy is conserved in the potential-flow simulations by rescaling the free-surface elevation and the surface potential at the end of every time step to generate new quantities.
\begin{eqnarray}
\eta^{\rm (NEW)} & = & S(t) \eta \nonumber \\
(\phi^s)^{\rm (NEW)} & = & S(t) \phi^s \;\; ,
\end{eqnarray}
\noindent where $S(t)$ is a scaling factor equal to the square root of the ratio of the current total energy to the initial total energy:
\begin{eqnarray}
S(t) =\left( \frac{E(t)}{E(0)} \right)^\frac{1}{2} \;\; .
\end{eqnarray}
As a result, energy is {\it pumped} into the free-surface waves.   Pumping nonlinear simulations of ocean waves  can be used to establish a $k^{-3}$ wavenumber dependence in wave spectra through the action of nonlinear wave interactions.    For example, pumped HOS simulations with third or higher order will fill in low-passed realizations of short-crested seas with a $k^{-3}$ power-law behavior corresponding to a saturated spectrum.  However, over long periods of time, energy will tend to pileup at high wavenumbers without an energy drain.

\subsubsection{Pumping for VOF method}

The total energy as a function of time is
\begin{equation}
E(t)=\int_V dV  \left(\frac{\rho(t)}{2}u_i(t)u_i(t) +  \frac{\left[\rho(t)-\rho(0)\right] z}{F_r^2} \right) ,
\end{equation}
where the first term on the right-hand side is the kinetic energy, and the second term is the potential energy.  The volume integrals are performed over the entire volume of the computational domain ($V$).
Similar to the potential-flow methods, the velocity and density are rescaled:
\begin{eqnarray}
u_i^{\rm (NEW)} & = & S(t) u_i \nonumber \\
\rho({\bf x},z_s,t)^{\rm (NEW)} & = & \rho({\bf x},z,t) \;\; ,
\end{eqnarray}
where $z_s = z+S(t) (z-z_o)$ is a stretching factor that increases the potential energy in proportion to the kinetic energy.  $z_o$ is a reference datum for the potential energy.   As before, the scaling factor $S(t)$ is equal to the square root of the ratio of the current total energy to the initial total energy.

\subsection{Free-Surface Adjustment}

Numerical simulations of nonlinear progressive waves are prone to developing spurious high-frequency standing waves unless the flow field is given sufficient time to adjust \citep{dommermuth00}.    An adjustment procedure is developed that allows nonlinear free-surface simulations to be initialized with linear solutions for HOS and FSM.  The adjustment scheme allows the natural  development of nonlinear self-wave (locked modes) and inter-wave (free modes) interactions.  Linear Airy waves are adjusted to generate nonlinear waves in HOS and FSM simulations.   The nonlinear terms in HOS and FSM are isolated and slowly activated.  In the case of NFA, nonlinear waves are generated from rest by applying a surface stress.    NFA requires a different approach than either HOS or FSM because the necessary two-phase solutions for progressive waves are not readily available.

\subsubsection{Adjustment for Spectral Method}

We assign an adjustment factor ($A(t)$) that slowly turns on nonlinearity as a function of time:
\begin{eqnarray}
A(t) = 1- \exp(-(\frac{t}{T_o})^2) \;\;   ,
\end{eqnarray}
where $T_o$ is the adjustment time.

For the spectral method, the following procedure is used to adjust the free-surface boundary conditions with no effects due to surface tension:
\begin{eqnarray}
\frac{\partial \phi^s}{\partial t} & = & - \frac{1}{F_r^2} \eta +  \frac{2}{Re} (\frac{\partial^2 \phi}{\partial x^2}+ \frac{\partial^2 \phi}{\partial y^2} ) \nonumber \\
& + & [ \frac{1}{2} \left(1+\nabla_x \eta \cdot \nabla_x \eta \right) \frac{\partial \phi}{\partial z}^2  \nonumber \\
& - &  \frac{1}{2} \nabla_x \phi^s \cdot \nabla_x \phi^s ] A(t) \;\; {\rm on \; z= \eta} \nonumber \\
\frac{\partial \eta}{\partial t} & = &   W^{(1)}  + \frac{2}{Re} (\frac{\partial^2 \eta}{\partial x^2} + \frac{\partial^2 \eta}{\partial y^2} ) \nonumber \\
& + & [  (1+ \nabla_x \eta \cdot \nabla_x \eta )  \frac{\partial \phi}{\partial z}- W^{(1)}  \nonumber \\
& - &    \nabla_x \phi^s \cdot \nabla_x \eta ] A(t) \nonumber \\ & &  {\rm on \; z= \eta} \;\; ,
\end{eqnarray}
\noindent where $W^{(1)}$ is the leading-order component of the vertical velocity evaluated on the plane $z=0$.
\begin{eqnarray}
W^{(1)}= \frac{\partial \phi}{\partial z}^{(1)}|_{z=0}
\end{eqnarray}

\subsubsection{Adjustment for Mapping Method}

The free-surface mapping is slowly turned on to its full value using the following procedure:
\begin{eqnarray}
\zeta = \frac{z+h}{h+A(t) \eta(x,y)}  \;\; .
\end{eqnarray}
\noindent Corresponding to this mapping, the boundary-value problem is solved for $-h \leq z \leq A(t) \eta(x,y)$.   Similar to the HOS method, the free-surface boundary conditions are adjusted:
\begin{eqnarray}
\frac{\partial \phi^s}{\partial t} & = & - \frac{1}{F_r^2} \eta  + \frac{2}{Re} (\frac{\partial^2 \phi}{\partial x^2}+ \frac{\partial^2 \phi}{\partial y^2} ) + \frac{1}{W_e} (\eta_{xx}+\eta_{yy})  \nonumber \\
& + & [ \frac{1}{2} \left(1+\nabla_x \eta \cdot \nabla_x \eta \right) \frac{\partial \phi}{\partial z}^2  -  \frac{1}{2} \nabla_x \phi^s \cdot \nabla_x \phi^s \nonumber \\
& - &  \frac{1}{W_e} ( \nabla \cdot n + \eta_{xx}+\eta_{yy}) ] A(t) \;\; {\rm on \; z= A(t) \eta} \nonumber \\
\frac{\partial \eta}{\partial t} & = &   \frac{\partial \phi}{\partial z}  + \frac{2}{Re} (\frac{\partial^2 \eta}{\partial x^2} + \frac{\partial^2 \eta}{\partial y^2} ) \nonumber \\
& + & \left[ \nabla_x \eta \cdot \nabla_x \eta  \, \frac{\partial \phi}{\partial z} - \nabla_x \phi^s \cdot \nabla_x \eta \right] A(t) \nonumber \\ & &  {\rm on \; z=A(t) \eta} 
\end{eqnarray}

\subsubsection{Adjustment for VOF Method}

The surface stress ($p_s$) in Equation (\ref{momentum}) can be used to apply a pressure to the interface to generate a known disturbance.  The formulation in terms of the gradient of Heaviside function ensures that the stress is applied only at the the free surface.   The stress is applied for a finite amount of time with an amplitude that is slowly ramped up and down to minimize transients.  The surface stress can be used to generate a linear superposition of waves in the following manner: 
\begin{equation}
p_s=G(t) \left[
\sum_{n=1}^N A_n \cos\left(k_n ( x+U_c t)-\omega_n (t-T_u) \right) \right]  \, ,
\label{eq:pa1}
\end{equation}
where $A_n$, $k_n$, and $\omega_n$ are respectively the Fourier amplitude, wavenumber, and frequency.  Typically, the wavenumber and wave frequency satisfy a linear dispersion relationship, $\omega_n^2= k_n/F^2_r$.   $U_c$ is the current velocity.   In a frame of reference that is fixed with the crest of the wave, $U_c$ equals the phase speed ($\omega_n/k_n$).  $T_u$ is an unwinding time that can be used to generate steep events at $t=T_u$.   $G(t)$ ramps up and down the stress for $0 \leq t \leq T_f$:
\begin{eqnarray}
G(t) =\frac{1}{2}\left[1-\cos\left(\frac{2 \pi t}{T_f}\right)\right]  \; .
\label{eq:pa2}
\end{eqnarray}
$G(t)=0$ for $t > T_f$.

\subsection{Seaway Representation}

A Pierson-Moskowitz spectrum is used to initialize simulations of a seaway.    The wavelength at the peak of the spectrum ($L_o$) is used to normalize length scales.   The velocity scale is $U_o=\sqrt{g L_o}$.   Based on these choices for $L_o$ and $U_o$, the Froude number equals one ($F_r=1$).  In normalized variables, the one-dimensional wavenumber spectrum is 
\begin{eqnarray}
\label{pierson1}
S(k)=\frac{\alpha}{2 k^3} \exp\left( -6 \pi^2/k^2  \right) \;\; ,
\end{eqnarray}
\noindent where $\alpha=8.1 \times 10^{-3}$ is Phillips' constant.

A cosine spreading function is used to simulate a directional spectrum.
\begin{eqnarray}
\label{pierson2}
D(\theta)= \frac{1}{2 \sqrt{\pi}} \frac{\Gamma(s+1)}{\Gamma(s+\frac{1}{2})} \cos^{2 s}( \frac{1}{2} (\theta -\theta_o) )  \;\; ,
\end{eqnarray}
\noindent where $s$ controls the amount of spreading and $\theta_o$ is the primary direction of the waves.  $\Gamma$ is the Gamma function. 

\section{Results} \label{sec:results}

Figure (\ref{modes_hos}) illustrates free-surface adjustment for HOS.  The HOS simulation is initialized using a single Airy wave with an initial  wave steepness $H/\lambda=0.05$, where $H$ is the wave height and $\lambda$ is the wave length.   The simulation is two dimensional with  64 de-aliased Fourier modes.   An eighth-order approximation is used.  Ê The Froude number is one,  $F_r=1$.  The length and depth of the domain are respectively 1 and 0.5.  The simulation is performed for 4000 time steps with $\Delta t=0.05$.       The period of adjustment is $T_o=8$.   Smoothing is applied every 5 time steps with $\gamma_c=0.9$ (see Equation \ref{smooth_spectral}).   Figure (\ref{modes_hos}) compares HOS predictions to an exact Stokes wave solution up to the seventh harmonic.  There is some evidence of ringing that could be reduced by increasing the period of adjustment \citep{dommermuth00}.

Figure (\ref{modes_fsm}) illustrates free-surface adjustment for FSM.  The FSM simulation is initialized using a single Airy wave with an initial  wave steepness $H/\lambda=0.05$.   The simulation is two dimensional with  \mbox{$256 \times 129=33,024$} grid points.  Ê The Froude number is one,  $F_r=1$.  The length and depth of the domain are respectively 1 and 0.5.  The simulation is performed for 4000 time steps with $\Delta t=0.05$.       The period of adjustment is $T_o=8$.   Smoothing is not used.  Comparing Figure (\ref{modes_fsm}) to (\ref{modes_hos}) show that FSM has slightly less ringing than HOS with similar accuracy.

Figure (\ref{modes_nfa}) illustrates free-surface adjustment for NFA. The NFA simulation uses the surface stress formulation (see Equations \ref{eq:pa1} and \ref{eq:pa2}) to generate a nonlinear wave.  The Froude number is one,  $F_r=1$.  A single Fourier mode is used with $A_0=0.003353$, $K_0=1$, and $\omega_0=1$.  A fixed frame of reference is used with $U_c=-1$. The forcing period is  $T_f =12 \pi$.  The NFA simulation is two dimensional with \mbox{$4096 \times 2048=8,388,608$} grid points.   The length, depth of water, and height of air are respectively $2\pi$, $\pi$, and $\pi$.  The simulation is performed for 100,000 time steps with $\Delta t=0.0005$.  Figure~(\ref{modes_nfa}) shows that non-linearity is established without ringing or sloshing in the NFA simulation.   In addition, the bound harmonics are correctly predicted up to 6th order.  Some errors are evident at the highest harmonics due the piecewise reconstruction of the free-surface interface.   When this technique is generalized to a sum of Fourier modes, VOF formulations like NFA could be used to simulate a seaway.   The primary motivation is to enable direct computation of wave breaking without modeling similar to the computations of \citet{nfa1}.

\begin{figure}[h]
\begin{center}
\includegraphics[width=\linewidth]{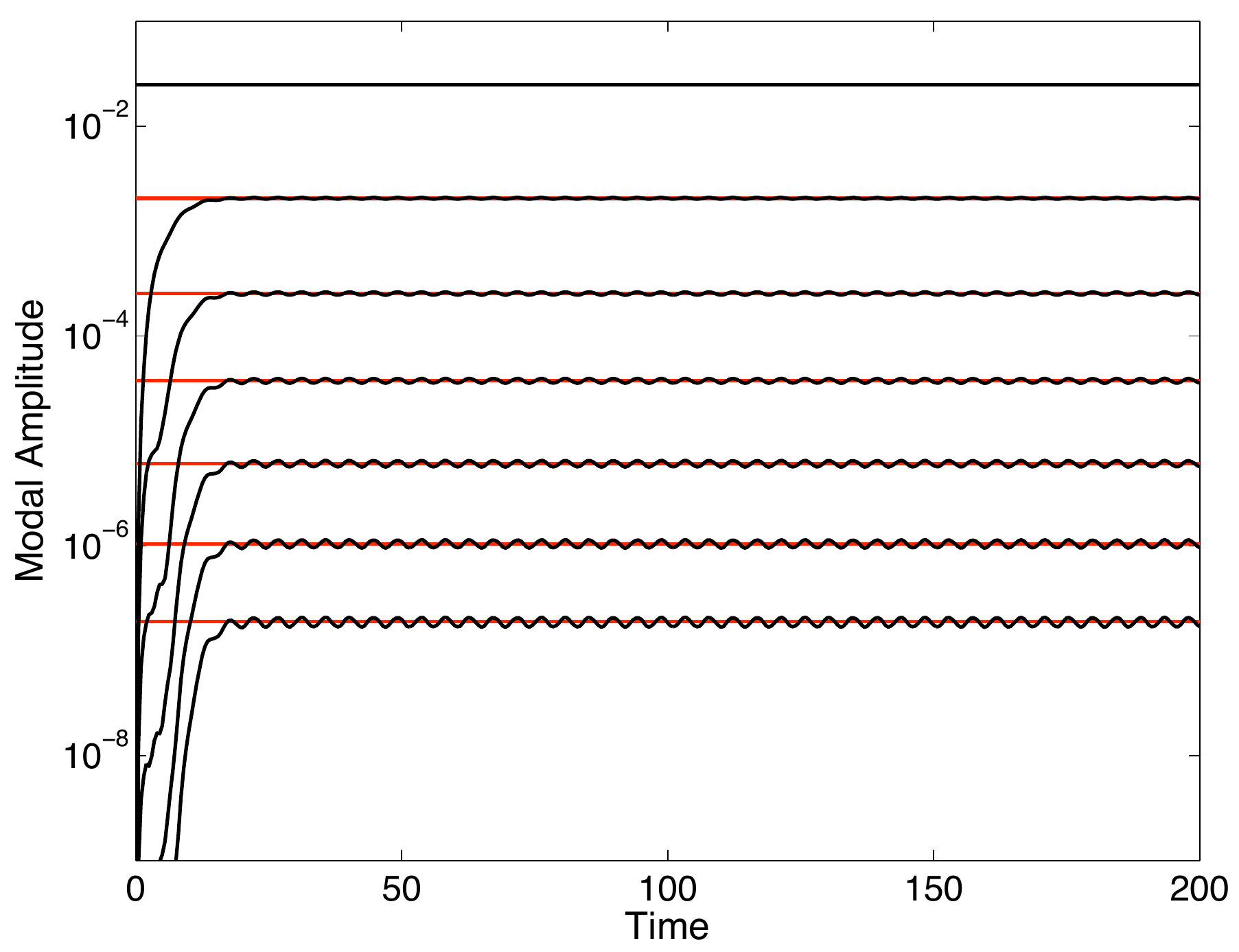}
\caption{\label{modes_hos}  Adjusted Stokes wave simulation for HOS.  Exact Stokes wave: \curvered. HOS: \curve. Modal amplitudes are plotted, 1-7.}
\end{center}
\end{figure}

\begin{figure}[h]
\begin{center}
\includegraphics[width=\linewidth]{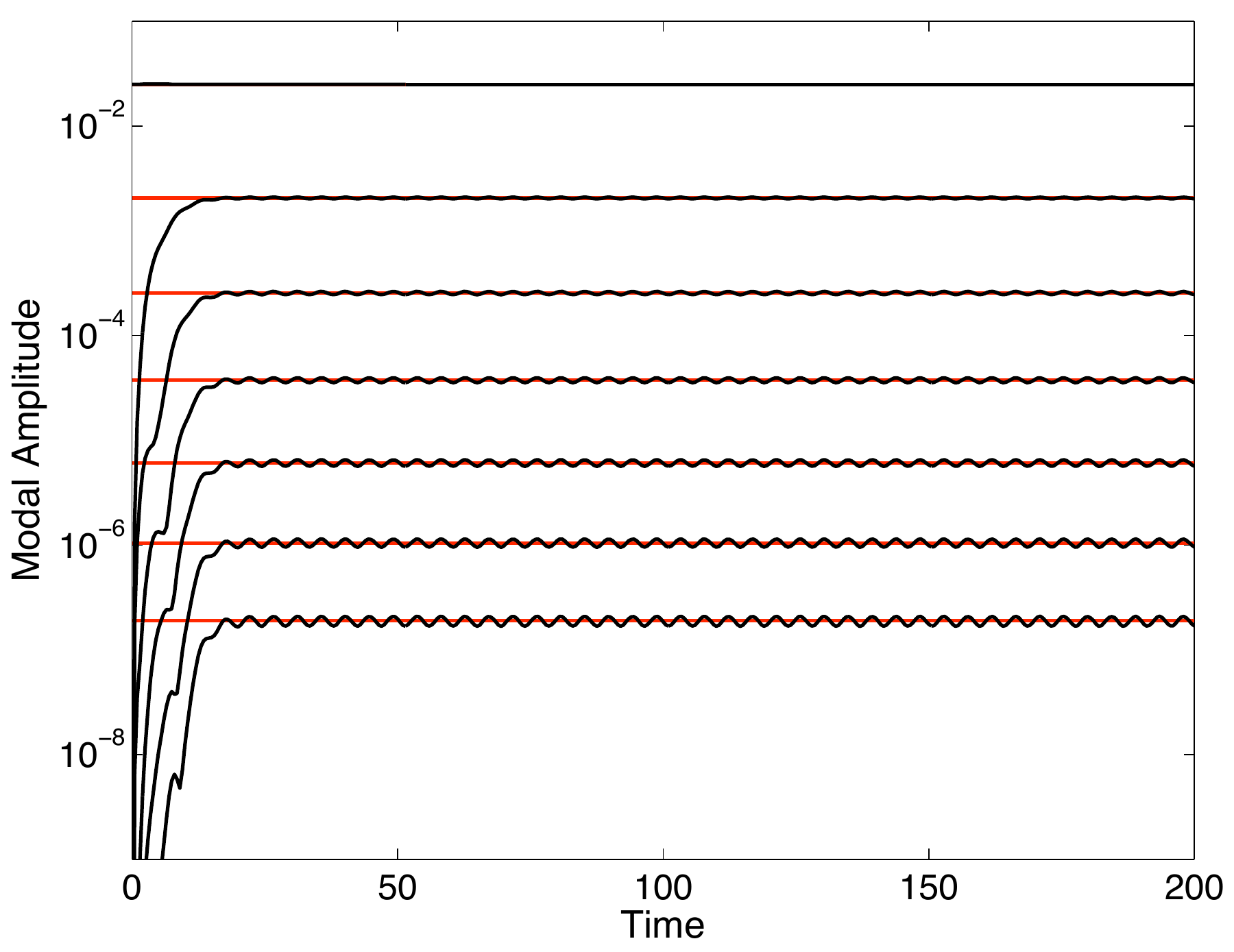}
\caption{\label{modes_fsm}  Adjusted Stokes wave simulation for FSM.  Exact Stokes wave: \curvered. FSM: \curve.  Modal amplitudes are plotted, 1-7.}
\end{center}
\end{figure}

\begin{figure}[h]
\begin{center}
\includegraphics[width=\linewidth]{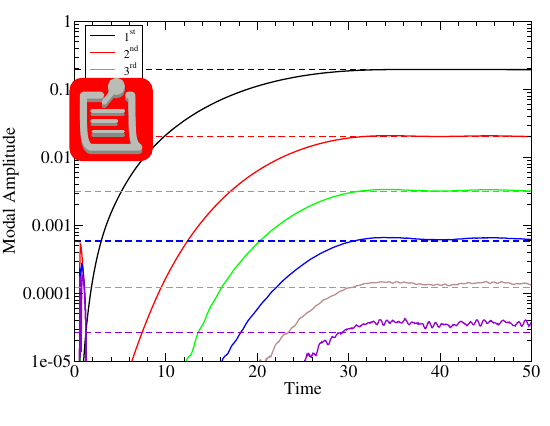}
\caption{\label{modes_nfa}  Generation of a Stokes wave for NFA. Dashed lines denote an exact Stokes wave.  Solid lines denote NFA.}
\end{center}
\end{figure}

The capabilities of HOS and FSM are assessed by simulating short waves riding on long waves.    \citet{longuet60} derive wave-action equations that show that short waves steepen at the crest of the long wave and flatten near the trough.      \citet{zhang93} show that HOS converges slowly in the case of short waves riding on long waves and may even diverge if truncated at finite order.   Accurate modeling of the modulation of short waves by long waves is key to simulating broad-banded wave spectra.

For the numerical simulations of short waves riding on a long wave, the Froude number is $F_r=1$.   The simulations are two dimensional.  The length and water depth are respectively 1 and 1/2.   HOS uses a fourth-order approximation.  HOS uses smoothing every 5 time steps with $\gamma_c=0.9$.   FSM does not use smoothing.  The adjustment time for both simulations is $T_o=8$.  The height to wave length ratios for the long and short waves are respectively $H_1/\lambda_1=0.025$ and  $H_2/\lambda_2=0.025$. Table~(\ref{shortwavetable}) shows additional details for each wave study for HOS and FSM.   $\lambda_1/\lambda_2$ is the wavelength ratio.   $N_x$ is the number of de-aliased Fourier modes for HOS.  $N_x \times N_z$ is the number of points used in FSM.   $F_z$ is the amount of grid stretching that is used in FSM near the free surface.    $N_t$ and $\Delta t$ are the number of time steps and the time step for both HOS and FSM.  For each wave-length ratio, one simulation is performed with the short wave riding on the long wave, and another simulation is performed with just the long wave at the same resolution.   The results of the two simulations are differenced to measure the interaction between the long wave and the short wave.  Direct comparisons between HOS and FSM are difficult because the adjustment procedures between the two methods are slightly different.

HOS and FSM are comparable for $\lambda_1/\lambda_2 \leq 16$, but for $\lambda_1/\lambda_2 = 32$, HOS breaks down.    Figure~\ref{shortwavefigure} compares HOS and FSM just before the HOS simulation broke down.  The long wave disturbances have been subtracted out, but there are still small residuals at low wave numbers due to nonlinear wave interactions.  FSM runs two and a half times longer without breaking down.  FSM is capable of simulating a wider bandwidth than HOS, but neither HOS or FSM can directly simulate wave breaking due to single-valued approximation.

\begin{table*}[h]
\begin{center}
\begin{tabular}{||c||c|c|c||c|c|c|c||} \hline
                      & \multicolumn{3}{|c||}{HOS} & \multicolumn{4}{c||}{FSM} \\ \hline
$\lambda_1/\lambda_2$ & $N_x$ & $N_t$ & $\Delta t$ & $N_x \times N_z$ & $F_z $ & $N_t$ & $\Delta t$  \\ \hline
2    & 64  & 4000 & 0.05 & $256 \times 129$ & 1 &  4,000 & 0.05 \\ \hline
4    & 128 & 8000 & 0.025 & $256 \times 129$ & 1 &  4,000 & 0.05 \\ \hline
8    & 128 & 8000 & 0.025 & $512 \times 129$ & 2 & 8,000 & 0.025 \\ \hline
16  & 256 & 8000 & 0.025 & $1024 \times 129$ & 4 & 16,000 & 0.0125 \\ \hline
32  & 512 & 16000 & 0.0125 & $2048 \times 257$ & 4 & 16,000 & 0.0125 \\ \hline
\end{tabular}
\caption{\label{shortwavetable} Numerical details of a short waves riding on a long wave.}
\end{center}
\end{table*}

\begin{figure*}
\begin{center}
\begin{tabular}{cc}
 HOS & FSM \\
\includegraphics[width=0.5\linewidth]{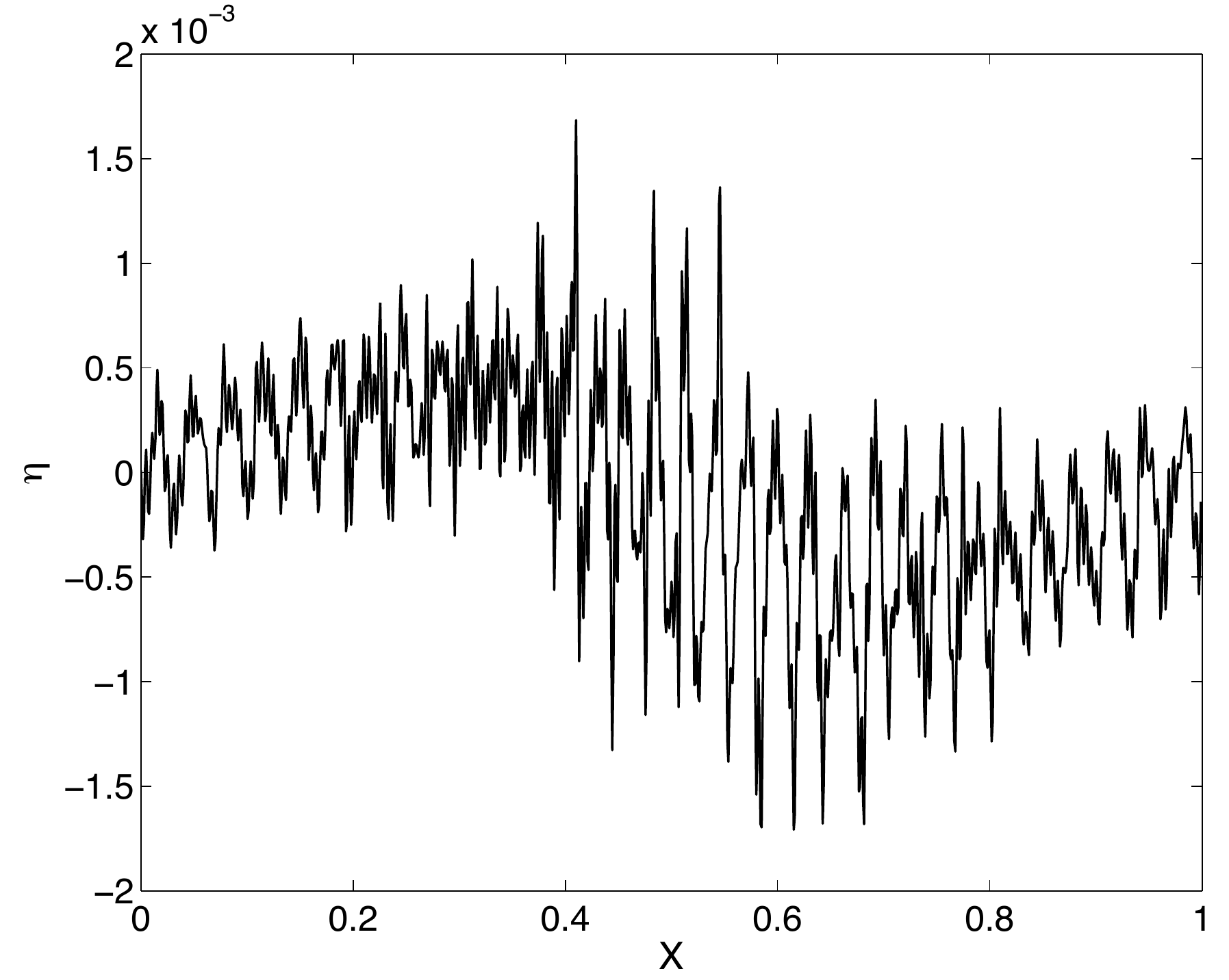}   & \includegraphics[width=0.5\linewidth]{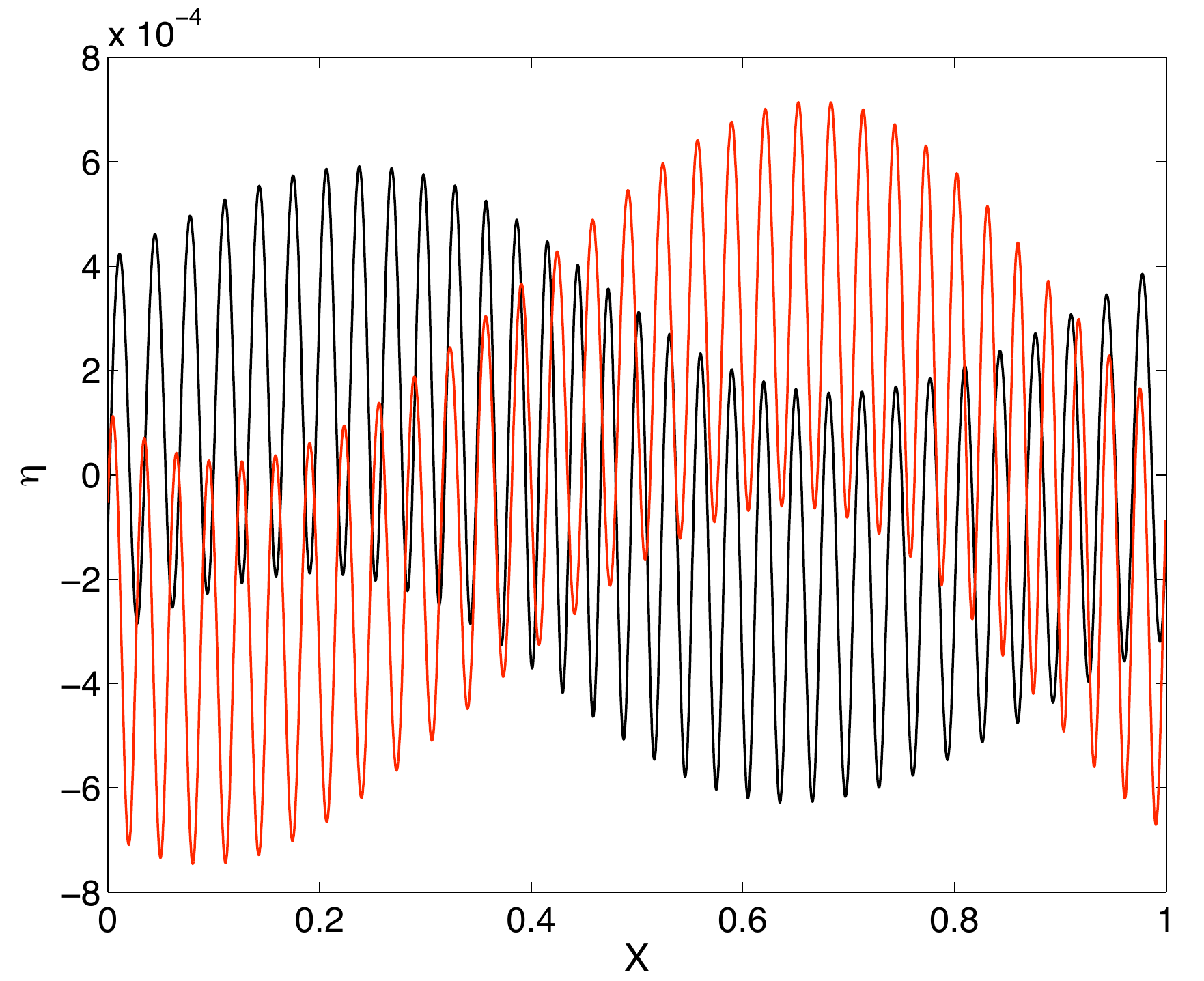}   \\ 
\end{tabular}
\end{center}
\caption{\label{shortwavefigure} Short wave modulation by a long wave.  Different vertical scales are used for (a) and (b).  (a) HOS, t=81.25.  (b) FSM, t=81.25: \curvered \, and t= 200: \curve. }
\end{figure*}

Simulations of short-crested seas are performed using HOS and FSM.  The simulations are initialized using a Pierson-Moskowitz spectrum with a slight angular spreading as specified by $s=10$ and $\theta_o=0$  (see Equations \ref{pierson1} and \ref{pierson2}) .    The Froude number is $F_r=1$.    The effects of surface tension and viscosity are not modeled.   The FSM simulation is three dimensional with $2048 \times 2048 \times 129= 541,065,216$ grid points.   The HOS simulation is three dimensional with $512 \times 512=262,144$ de-alaised Fourier modes.   The HOS uses a fourth-order approximation.  The length, width, and depth of the HOS and FSM simulations are respectively 20, 20, and 2 in normalized units.   As noted in the discussion of Equation~(\ref{pierson1}), all length scales are normalized by the length of the wave at the peak of the Pierson-Moskowitz spectrum. The vertical grid is geometrically stretched in the FSM simulation such that $\Delta z = 0.0078125$ at the free surface.   The FSM simulations are run for 17,000 time steps with $\Delta t=0.025$.   The HOS simulations are run for 40,000 time steps with $\Delta t=0.01$, but energy begins to pile up at high wave numbers after 10,000 time steps .   The period of adjustment for both HOS and FSM is $T_o=20$.    Smoothing is not initially used in the FSM simulation, but it is applied every 5 time steps after 10,000 time steps have elapsed.  Smoothing is used every time step throughout the HOS simulation with $\gamma_c =0.9$.   

Figure~\ref{statistics} compares skewness and kurtosis of the free-surface elevation for HOS and FSM.  On average, FSM has slightly more skewness and kurtosis than HOS.   Figure~\ref{spectrum} shows the free-surface spectra for HOS and FSM  for different time instants.   The results are time-averaged over 25 non-dimensional units of time.   The energy piles up at high wave numbers in the HOS simulation even though smoothing is applied every time step.    The FSM has a slight energy pile up, but turning on smoothing at $t=225$, eliminates the energy pileup.   FSM has more bandwidth than HOS with negligible energy pileup.   Figure~\ref{probability} shows probability distributions for surface elevation and u, v, and w velocities on the surface for HOS and FSM.  The u and v components of velocity are down-sea and cross-sea, respectively.  The results are time-averaged over 25 non-dimensional units of time.   As expected, the free-surface elevations are lower in the trough and higher in the crests for HOS and FSM compared to a Gaussian distribution.   The extreme waves crests are higher for FSM than HOS.    Comparing the free-surface probability distributions for $t=225$ (pre smoothing) and $t=425$ (post smoothing) for FSM, shows that smoothing tends to reduce extrema in the FSM simulation.  The extrema for the u-velocity probability distributions are also higher for FSM than HOS.   The FSM values are about 30\% higher.    However, smoothing again adversely affects extrema.   The same trends continue for the v and w-velocity probability distributions where FSM extrema are higher than HOS.

\begin{figure*}
\begin{center}
\begin{tabular}{cc}
 Skewness & Kurtosis \\
\includegraphics[width=0.5\linewidth]{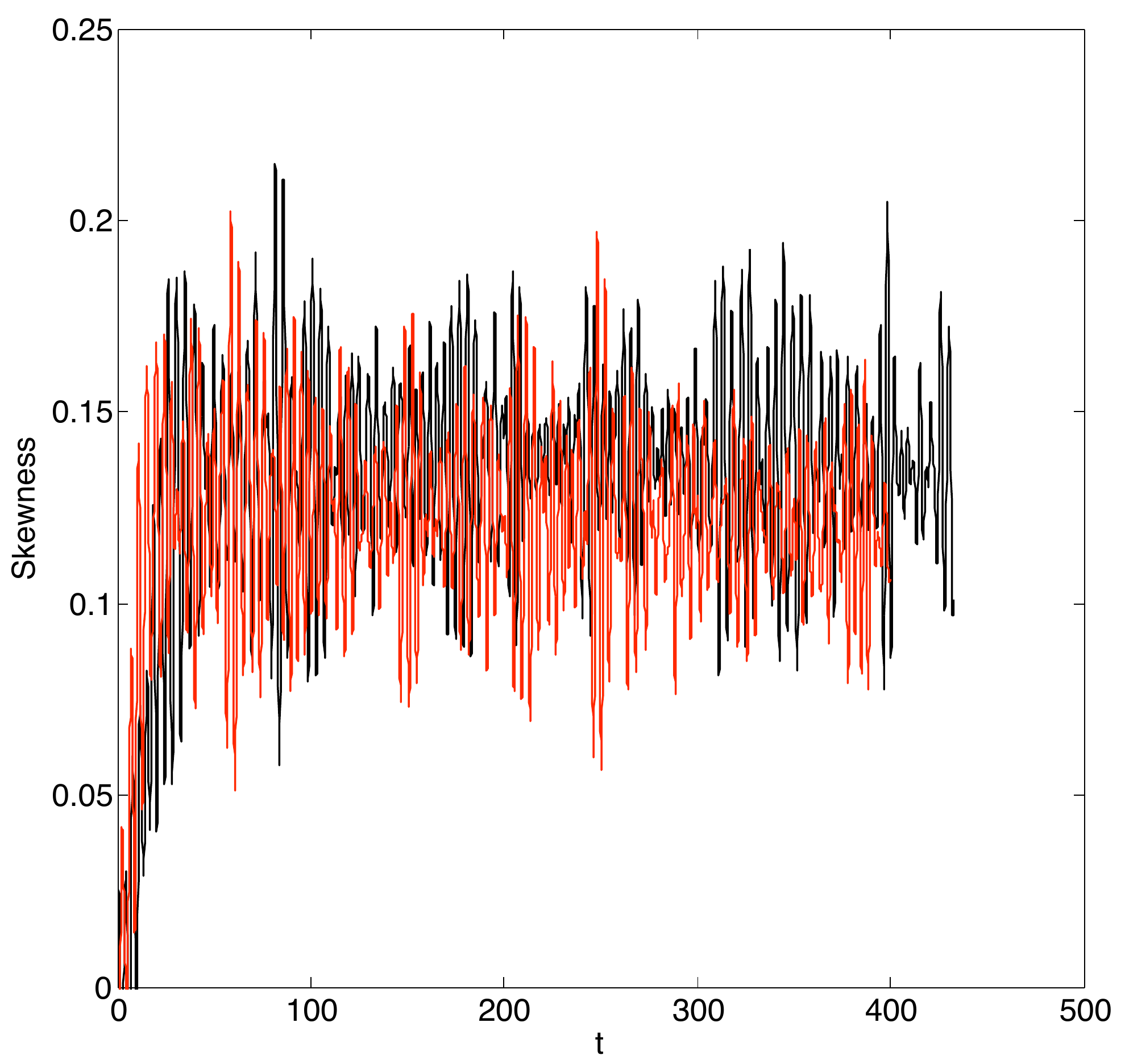}   & \includegraphics[width=0.5\linewidth]{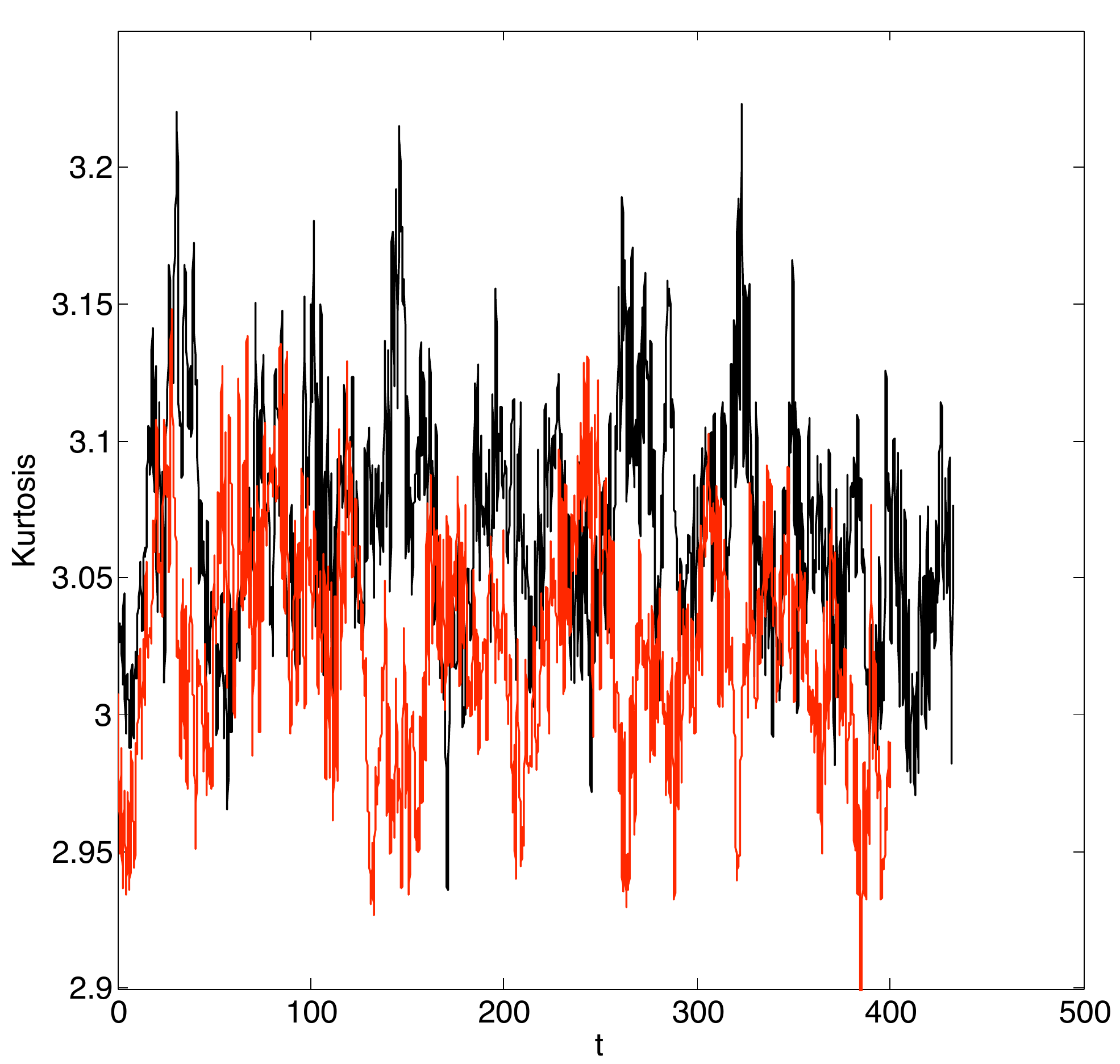}   \\ 
\end{tabular}
\end{center}
\caption{\label{statistics} Statistics of free-surface elevation.  HOS: \curvered. FSM: \curve.}
\end{figure*}

\begin{figure*}
\begin{center}
\begin{tabular}{cc}
 HOS & FSM \\
\includegraphics[width=0.5\linewidth]{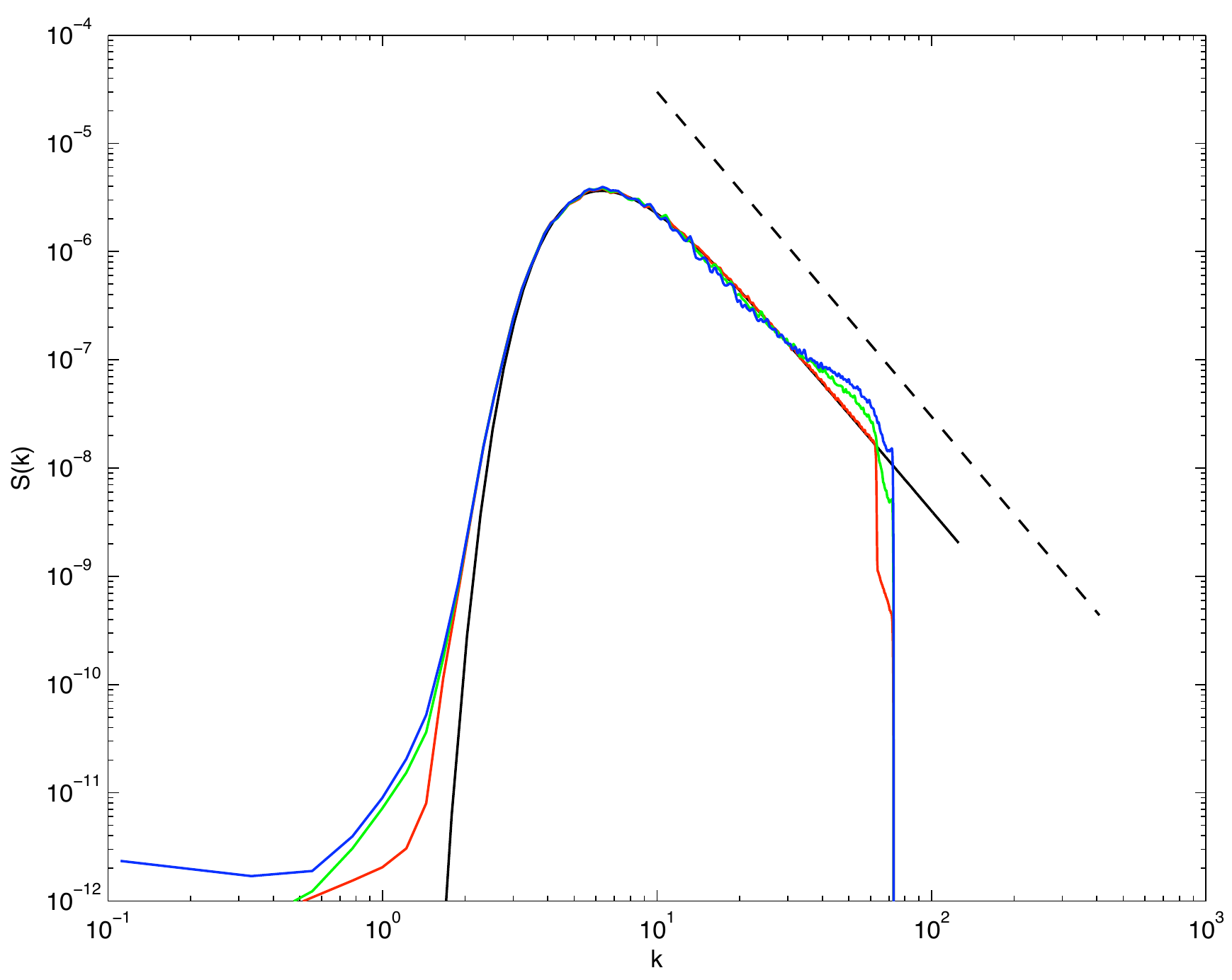}   & \includegraphics[width=0.5\linewidth]{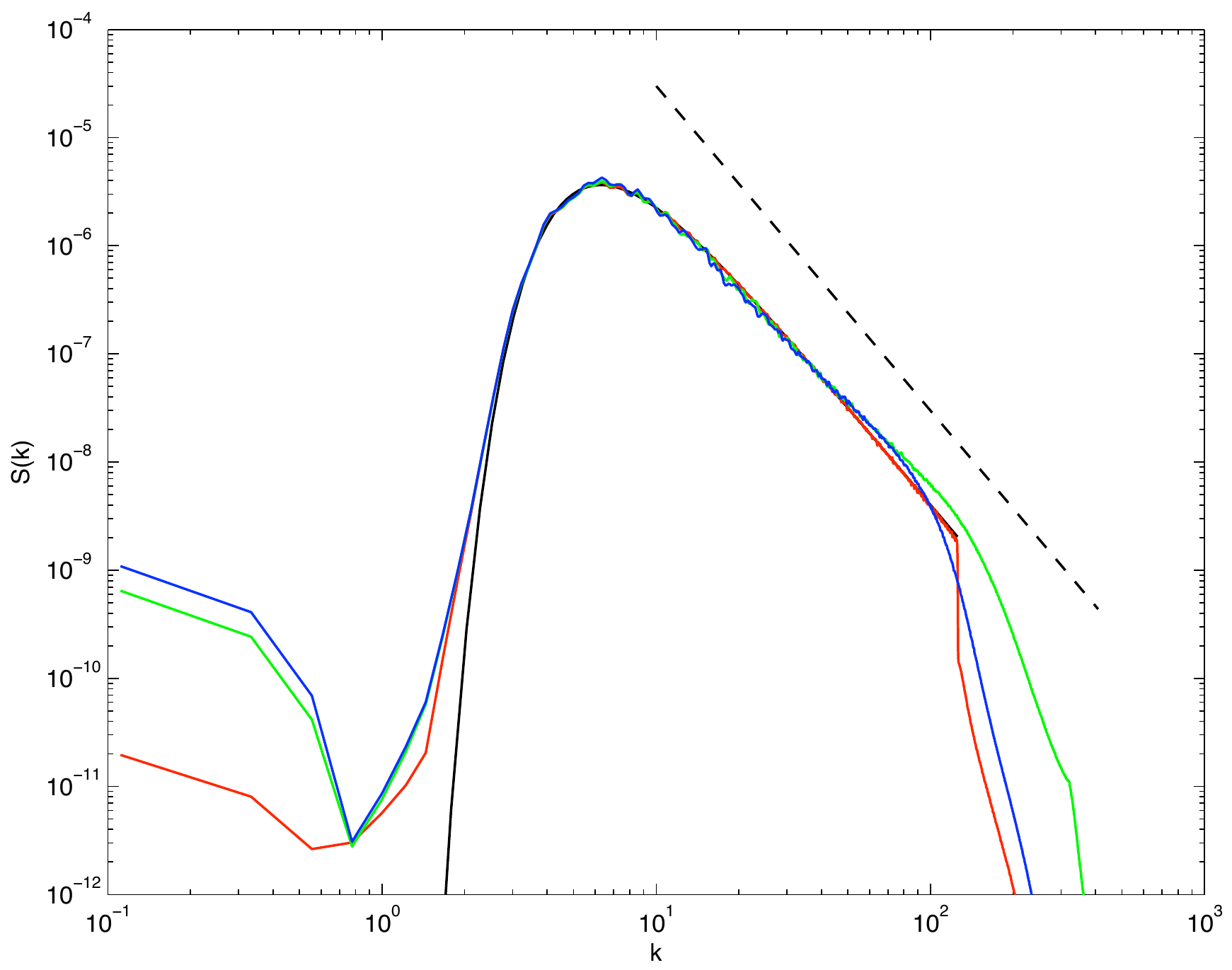}   \\ 
\end{tabular}
\end{center}
\caption{\label{spectrum} Free-surface spectra.  Pierson-Moskowitz spectrum: \curve.  $k^{-3}$: \curuud. t=25: \curvered. t=225: \curvegreen.  t= 425: \curveblue}
\end{figure*}

\begin{figure*}
\begin{center}
\begin{tabular}{ccc}
  & HOS & FSM \\
\raisebox{2.5cm}{A} & \includegraphics[width=0.3\linewidth]{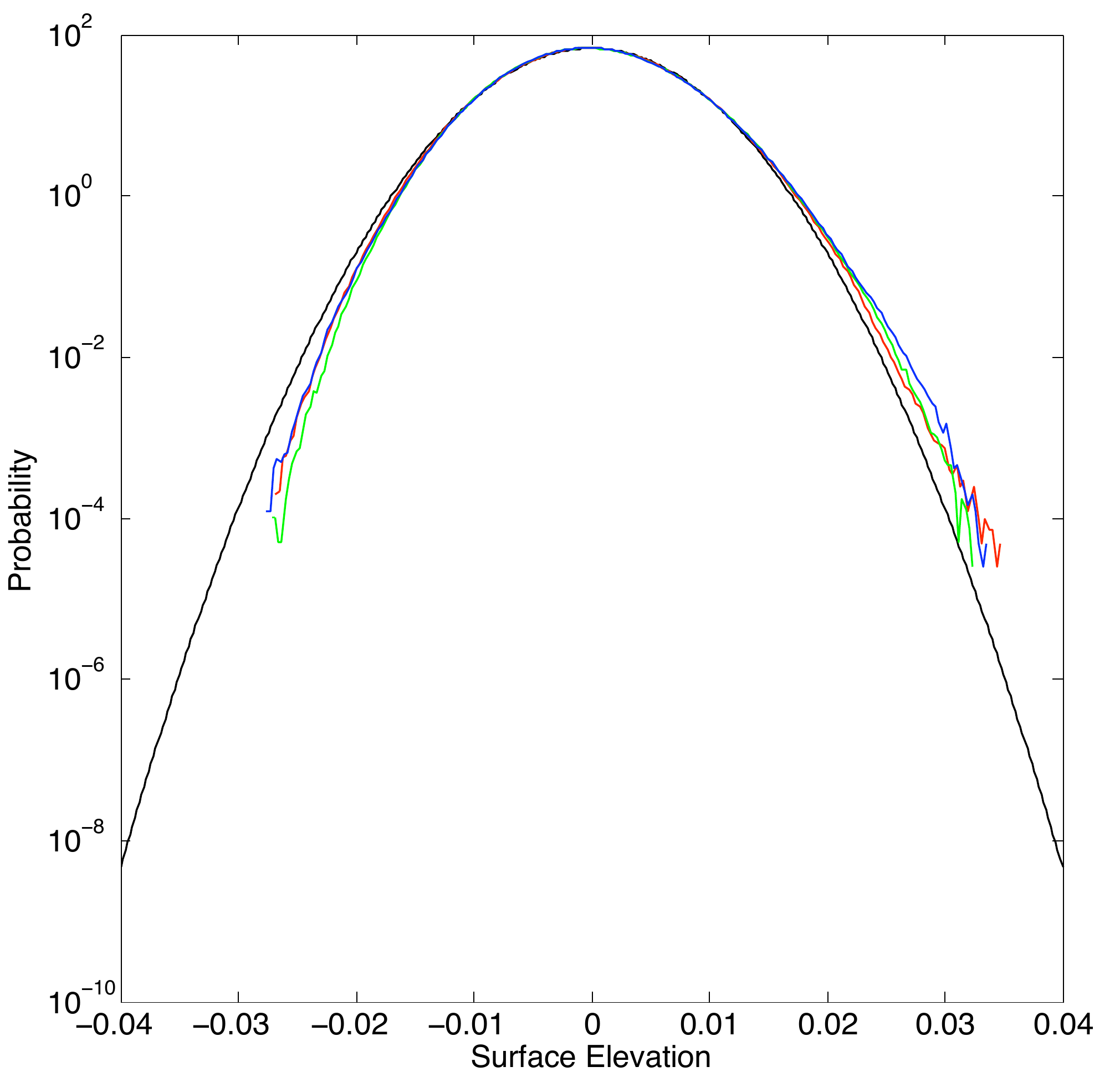}   & \includegraphics[width=0.3\linewidth]{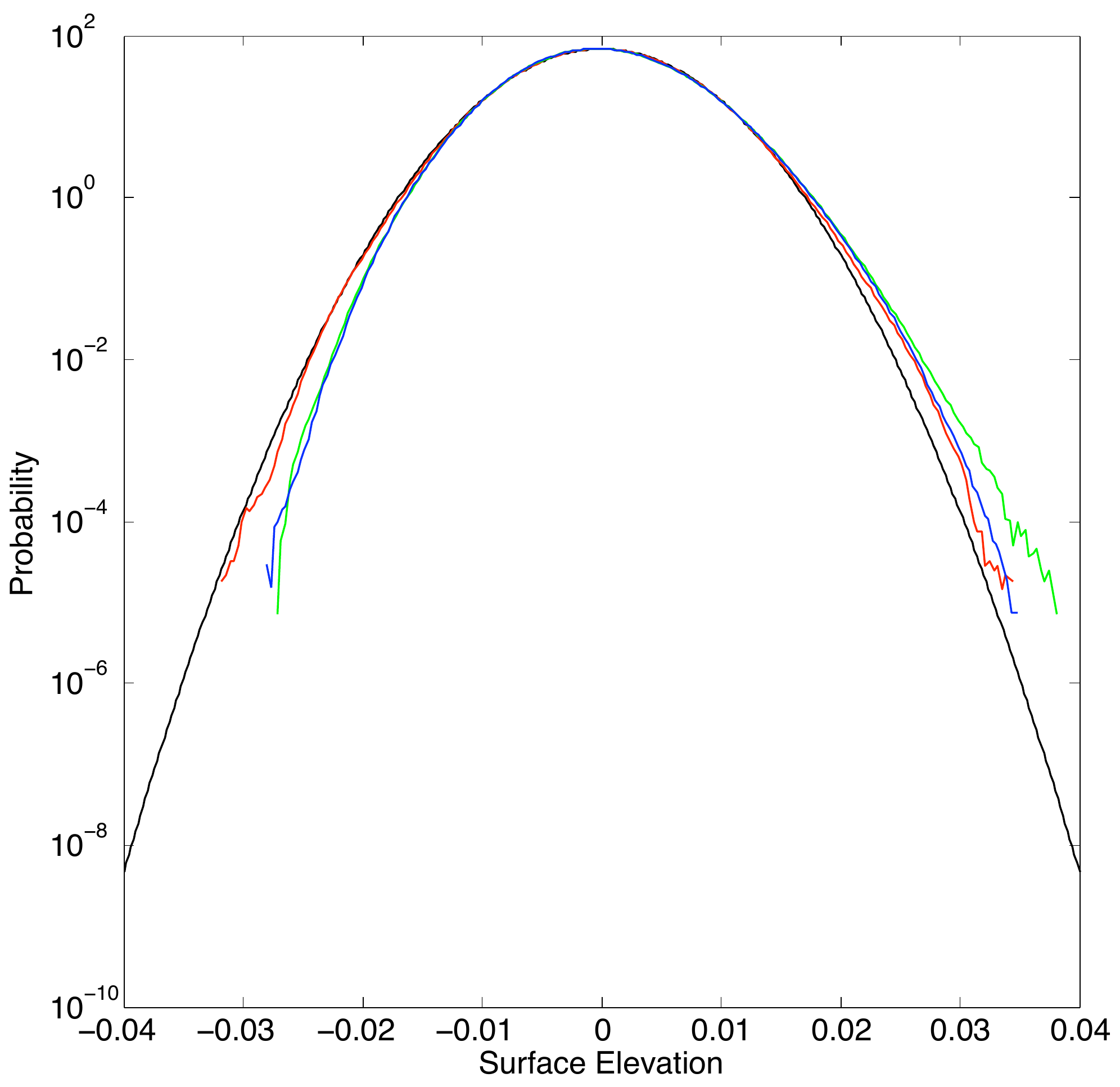}   \\ 
\raisebox{2.5cm}{B} & \includegraphics[width=0.3\linewidth]{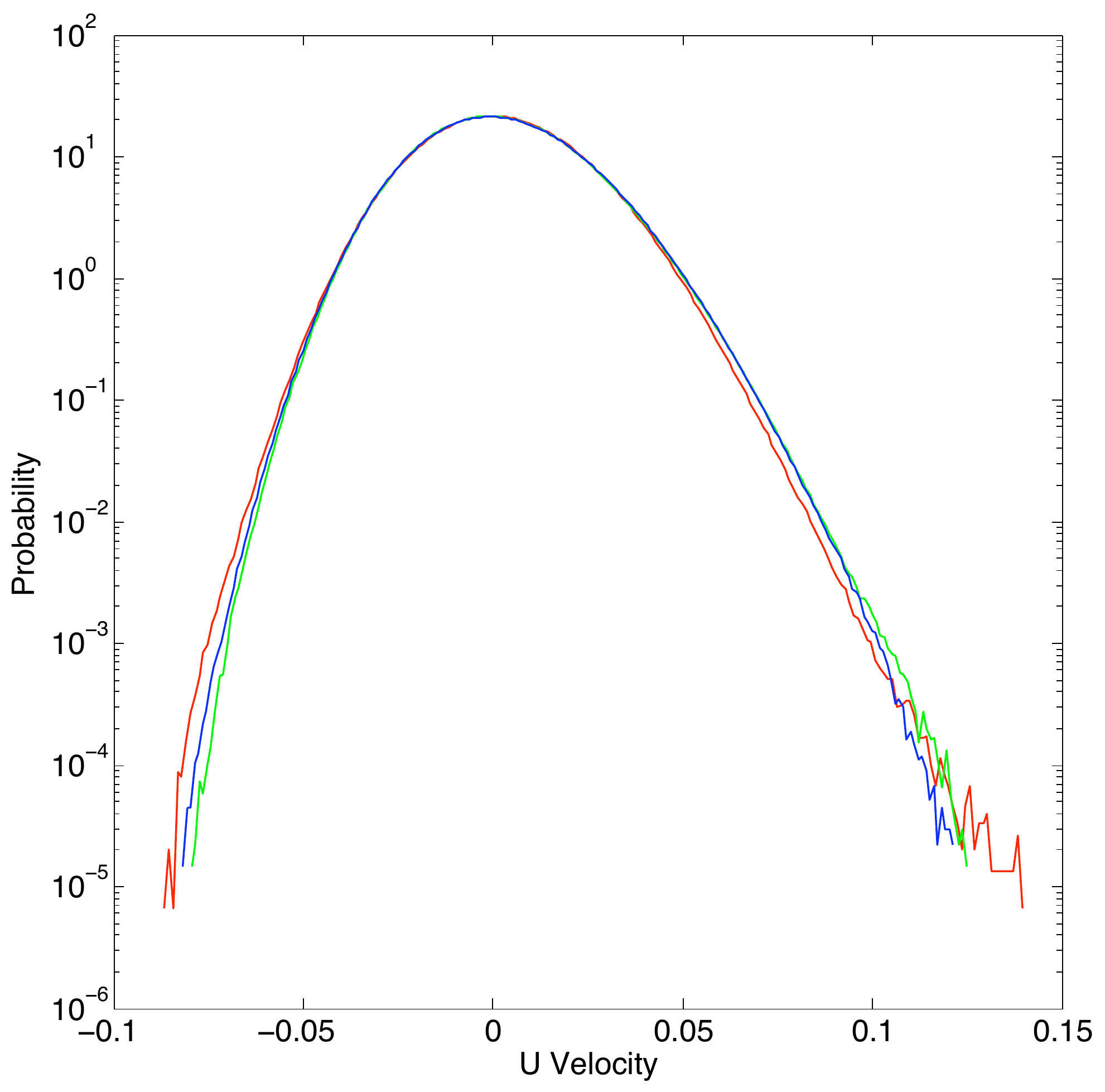}  & \includegraphics[width=0.3\linewidth]{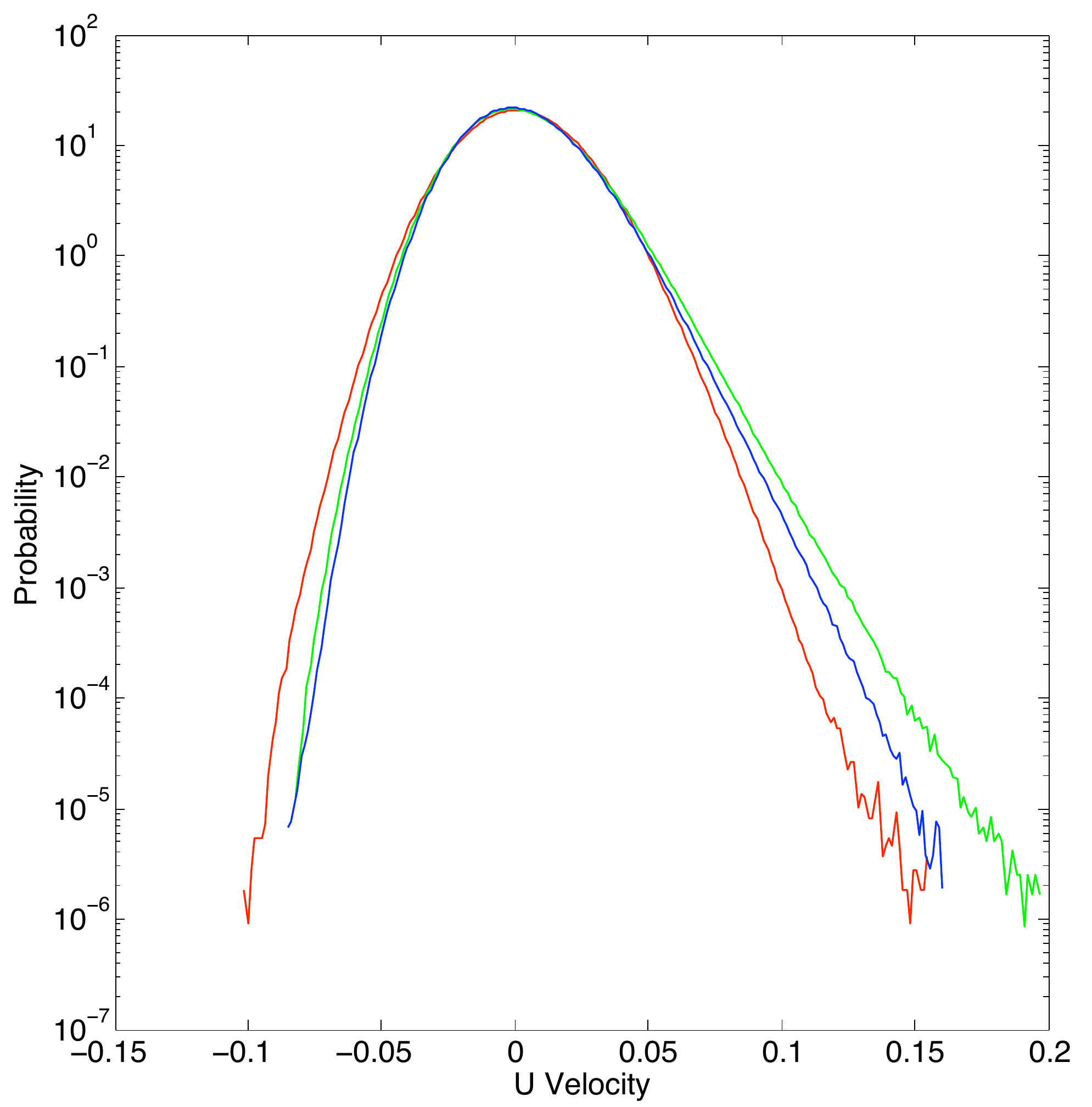}  \\ 
\raisebox{2.5cm}{C} & \includegraphics[width=0.3\linewidth]{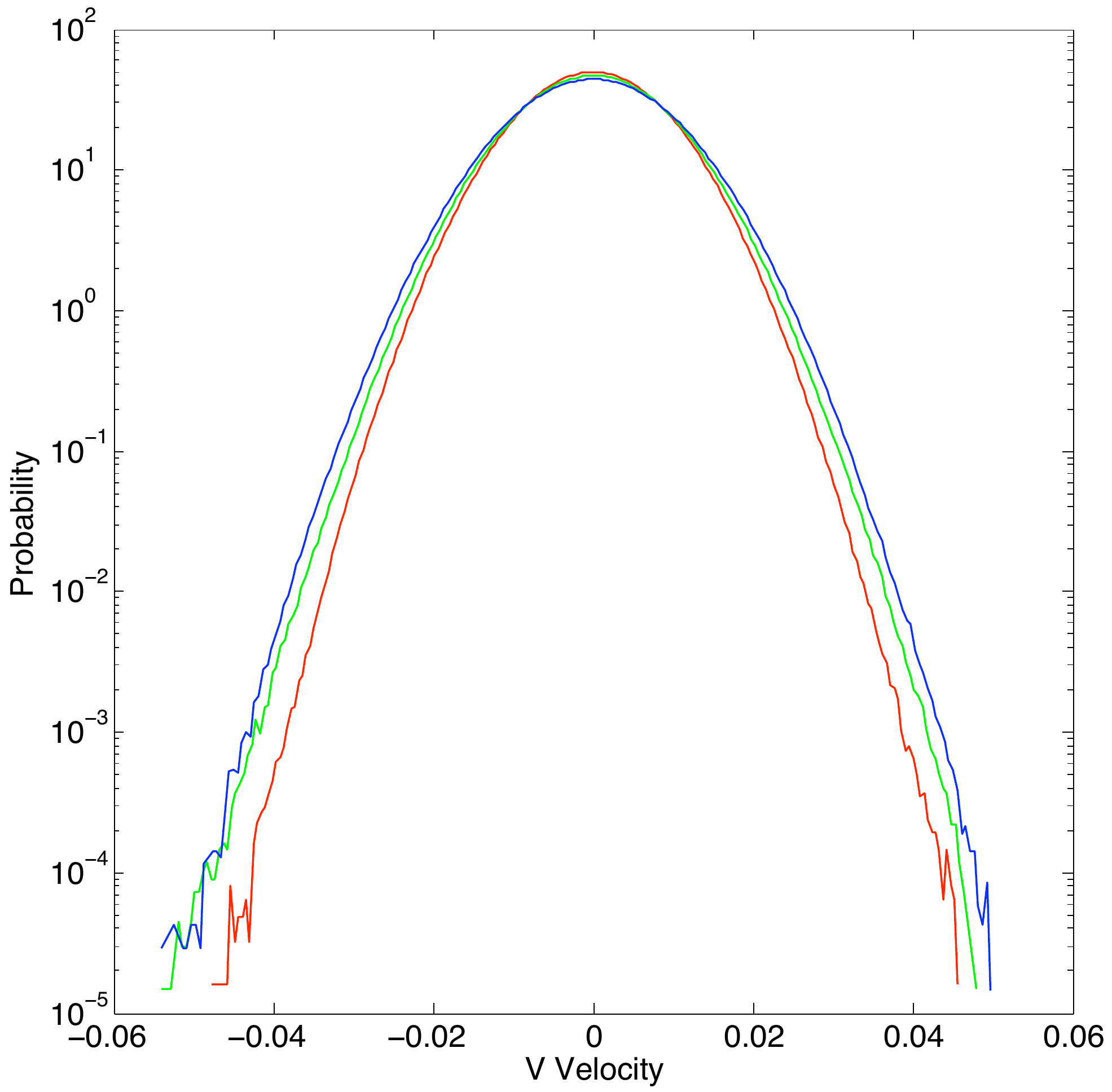}  & \includegraphics[width=0.3\linewidth]{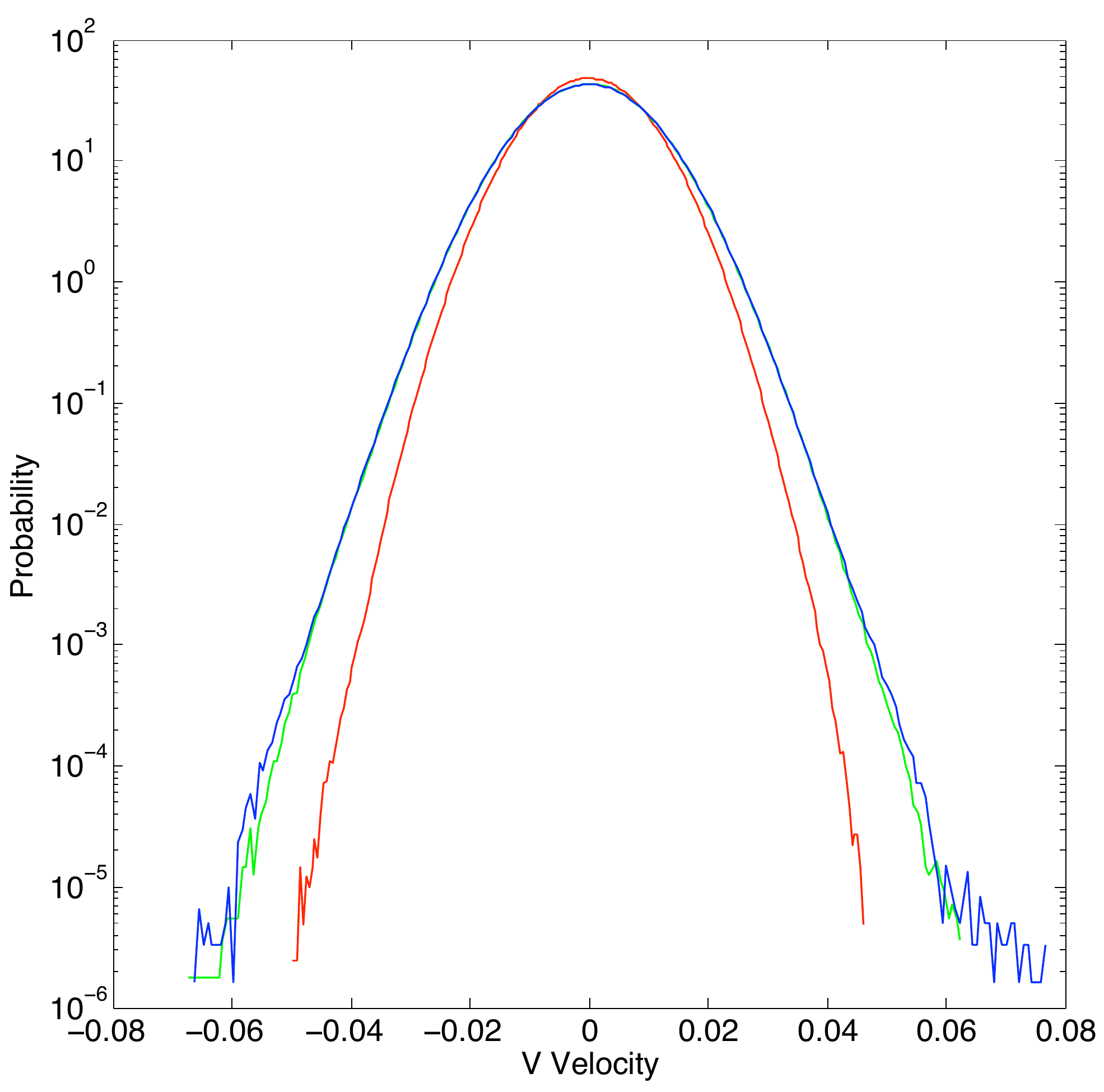}   \\ 
\raisebox{2.5cm}{D} & \includegraphics[width=0.3\linewidth]{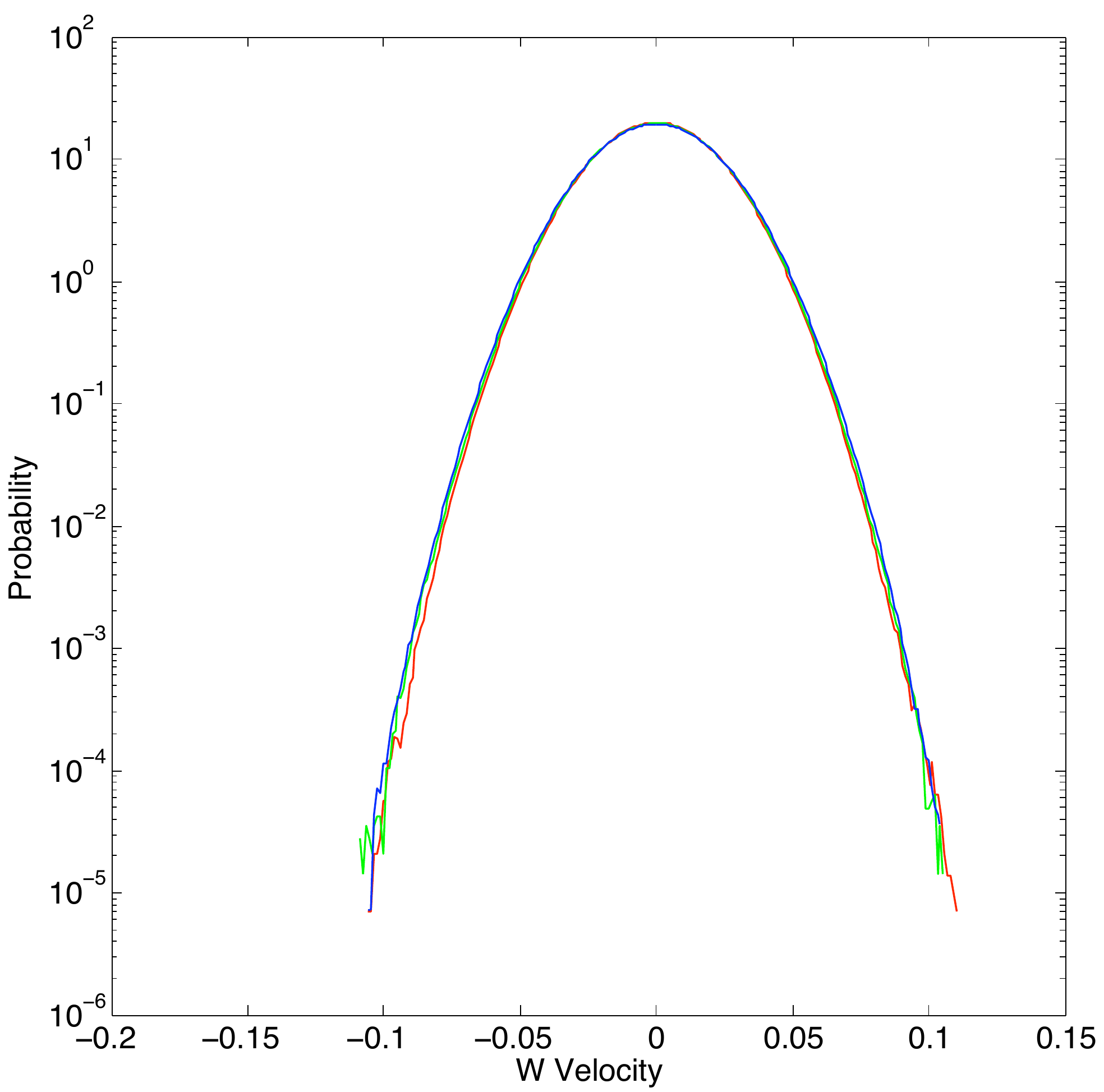} & \includegraphics[width=0.3\linewidth]{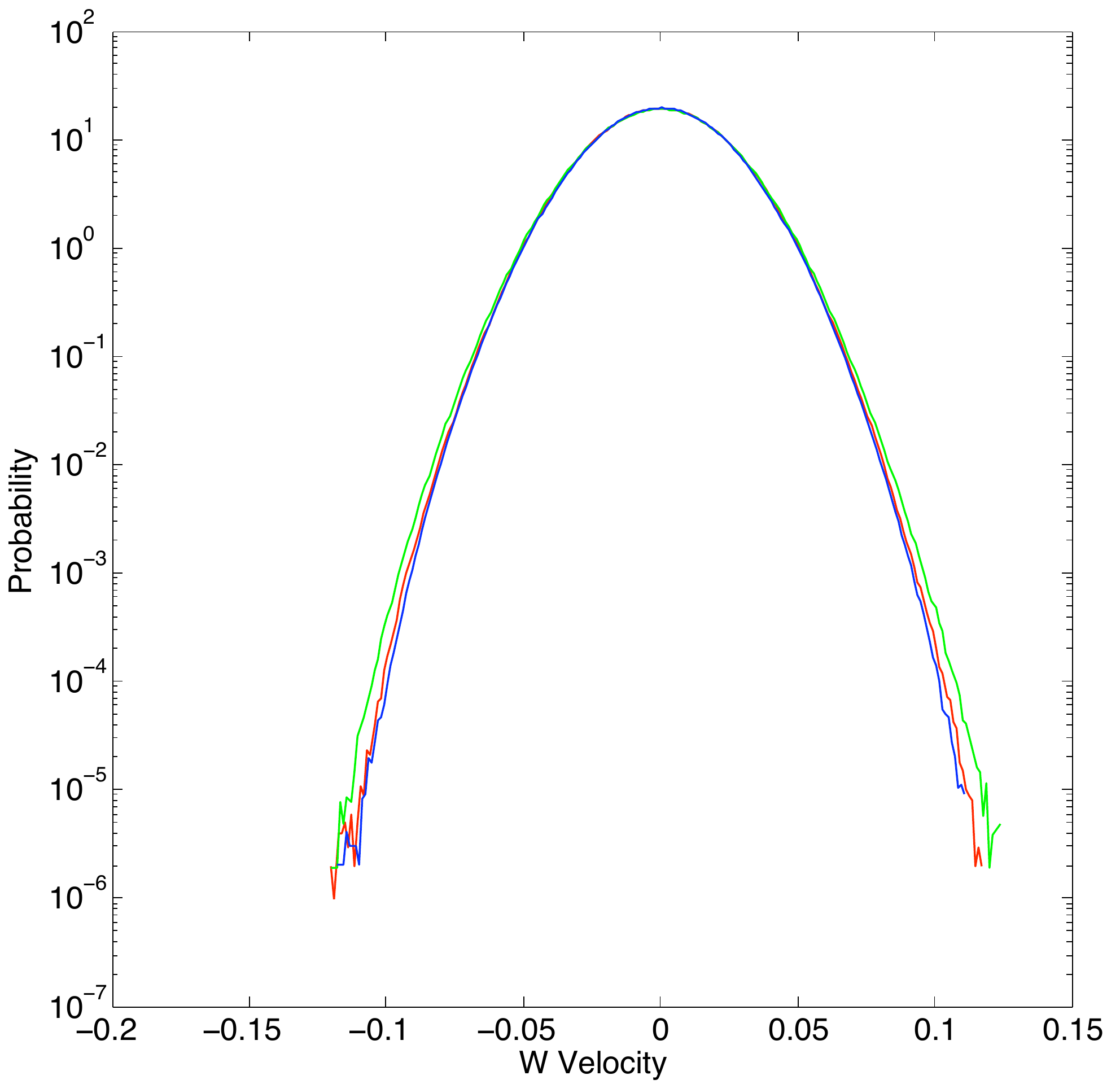}  \\
\end{tabular}
\end{center}
\caption{\label{probability} Probability distributions for FSM and HOS. Gaussain distribution: \curve.  t=25: \curvered. t=225: \curvegreen.  t= 425: \curveblue}
\end{figure*}

As an illustration of the FSM technique, we consider a numerical simulation of the formation of parasitic capillary waves.  We choose characteristics scales \mbox{$L_o=0.25m$}, \mbox{$U_o = \sqrt{g L_o} = 1.565 {\rm \; m/s}$}, and \mbox{$T_o = L_o/U_o = 0.1597 {\rm \; s}$}.   Based on these characteristic scales,  the non-dimensional parameters are \mbox{$F_r = U_o /\sqrt{g L_o} = 1$}, \mbox{$R_e = U_o L_o / \nu_w = 3.434 \times 10^5$}, and \mbox{$W_e = \rho_w U_o^2 L_o / \sigma_w = 8.283 \times 10^3$}, where \mbox{$g = 9.80665  {\rm \; m/s^2}$}, \mbox{$\rho_w=100 {\rm \; kg/m^3}$}, \mbox{$\nu_w = 1.14 \times 10^{-6}  {\rm \; m^2/s}$}, and  \mbox{$\sigma_w  = 0.074 {\; \rm N/m}$} .  The patch is 2m x 2m.   The grid resolution is 2048x2048.  The simulation is performed for 3,000 time steps with a non-dimensional time step \mbox{$\Delta t=0.005$}. The simulation is initially seeded with the tail end of a Pierson-Moskowitz wave spectrum just above the capillary regime with an initial wind speed equal to 10m/s.  The parasitic capillary waves form as a result of wave nonlinearity.  Figure (\ref{parasitic}a) shows the initial free-surface elevation.   Figure (\ref{parasitic}b) shows the simulation at a later time. The wave pattern forms crescent-shaped disturbances corresponding to catspaws.  It is interesting that these short waves form even without actually modeling the wind.   The waves form as a natural consequence of nonlinear wave interactions.   As has been observed in experiments, there is a hole in the spectral content between gravity-capillary waves and pure capillary waves.   This example illustrates that FSM can model very short waves riding on top of longer carrier waves, something that is not possible using HOS.

\begin{figure*}
\begin{center}
\begin{tabular}{cccc}
\raisebox{1cm}{(a)} & \includegraphics[width=0.45\linewidth]{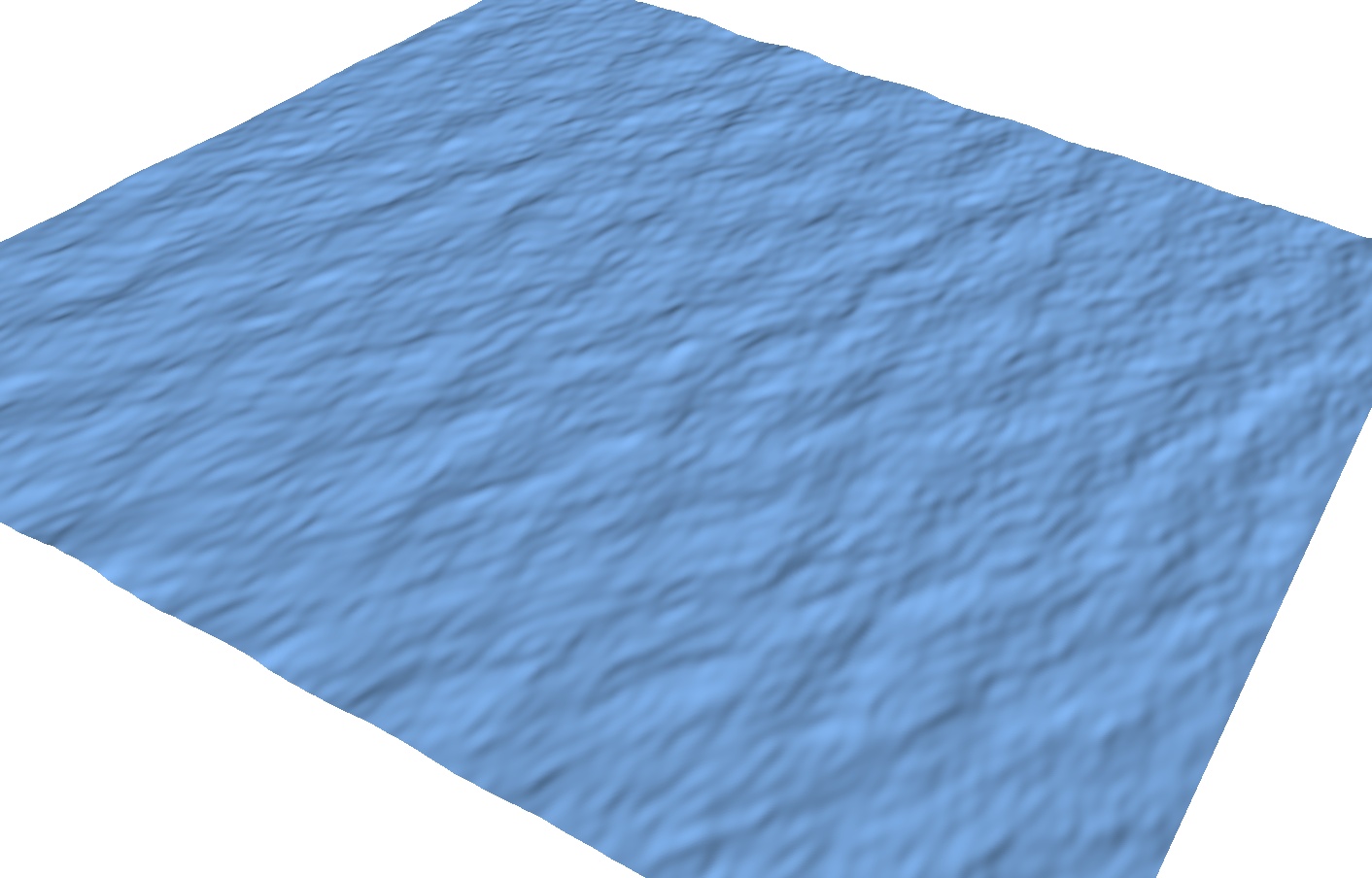}
\raisebox{1cm}{(b)} & \includegraphics[width=0.45\linewidth]{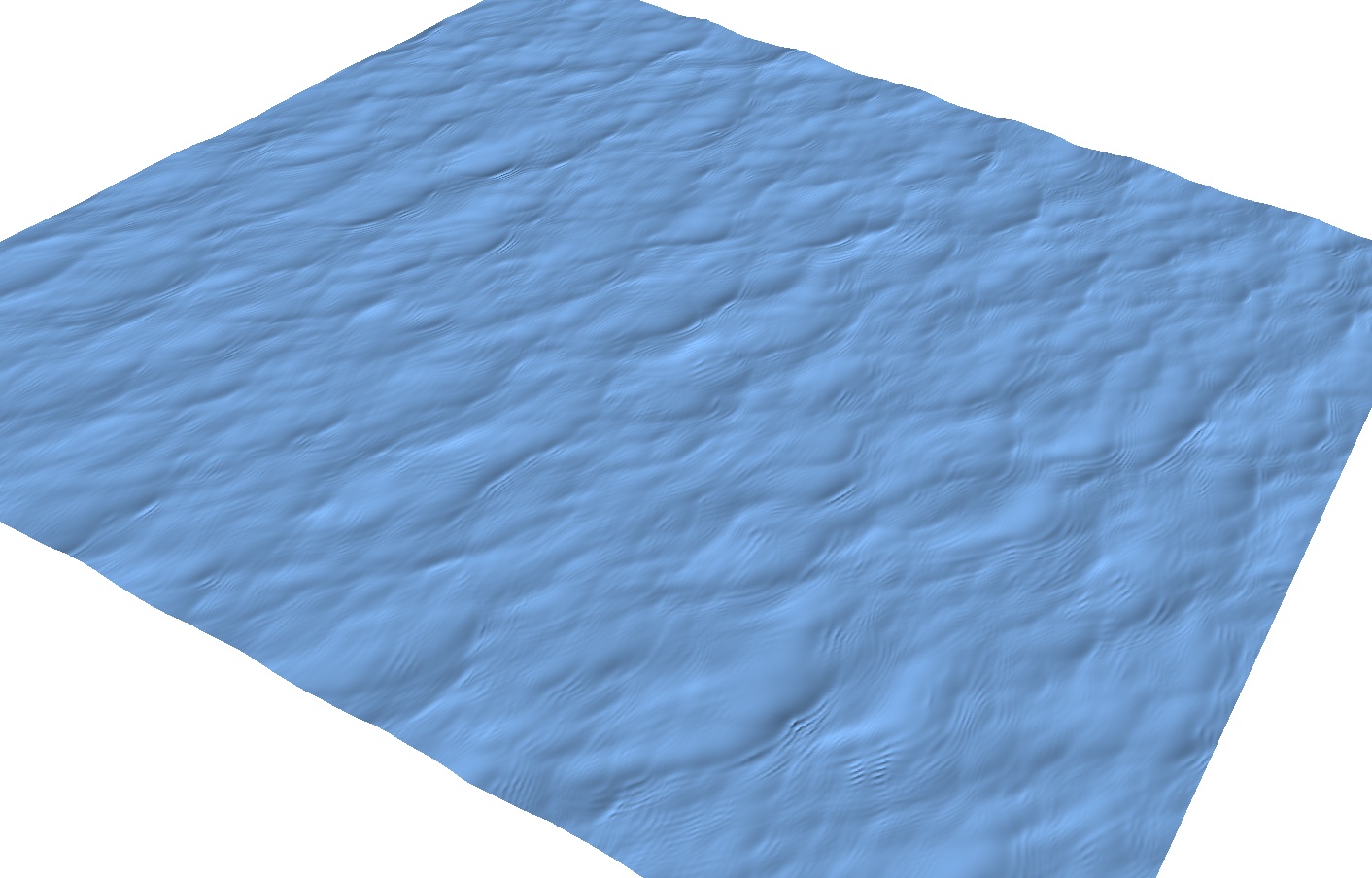}
\end{tabular}
\end{center}
\caption{\label{parasitic}  Formation of catspaws on an ocean surface.   The electronic version of this document can be zoomed in to see the catspaws more clearly.  (a) Free-surface elevation at t=0.  (b) Free-surface elevation at t=15.}
\end{figure*}

Figure (\ref{fission}) shows fissioning of a wave packet using HOS and NFA with comparisons to experiments.  The experiments are reported in \citet{su82}.  Details of the HOS simulation are available in \citet{dommermuth87}.  Su (1982) studied experimentally the evolution of wave groups that had initially square envelopes. For wave steepnesses ranging from 0.09 to 0.28, he measured the free-surface elevation at eight stations down the tank. For wave steepnesses greater than 0.14, he observed intense two-dimensional breaking at distances between ten and twenty carrier wavelengths from the wavemaker. Fifteen to twenty-five wavelengths from the wavemaker, crescent-shaped breaking waves often developed, and from twenty to forty-five wavelengths away, two-dimensional spilling breaking was common.  The initial packet has 5 waves.

For the NFA simulation, a wave packet is generated using a surface stress (see Equations \ref{eq:pa1} and \ref{eq:pa2}).  A single Fourier mode is used with $A_0=0.006$, $K_0=15$, and $\omega_0=\sqrt{15}$ with the Froude number, $F_r=1$.   A fixed frame of reference is used with $U_c=-\omega_0/(2k_0) \approx -0.12910$, which is the group velocity.  The forcing period is $T_f=5$.    The NFA simulation is two-dimensional with $4096 \times 256 =1,048,576$ grid points.  The length, depth, and height of the computational domain are $2 \pi$, $\pi/15$, and $\pi/15$. The simulation is performed for 120,000 time steps with $\Delta t=0.001$.  The air to water density ratio is $\rho_a/\rho_w=0.001205$.     Five-point density-weighted smoothing is applied every 1,000 time steps (see Equations \ref{smooth_velo}-\ref{project2}).   The stencil for the weighting function is (1/8,1/4,1/4,1/4,1/8), which is applied in sweeps along each cartesian direction.   The amplitude and duration of the surface stress are chosen to generate waves that match the initial wave steepness (0.15) in the experiments.  A tapering function $T(x)$ is applied to the surface stress in Equation (\ref{eq:pa1}) to generate a wave group.  The tapering function is specified in terms of tanh functions:
\begin{equation}
T(x)=\frac{1}{2}(\tanh(\sigma(x-x_b))-\tanh(\sigma(x-x_e)))
\end{equation} 
The parameter $\sigma=30/\pi$ controls the steepness of the wave packet, and $x_b=-\pi/3$ and $x_e=\pi/3$ determine the length of the wave packet.  Unlike the HOS simulations, the NFA simulations allow breaking.   For these NFA simulations, the breaking manifests itself as coaming at the crests of the waves at the front of the wave group.  At about forty wavelengths from the wavemaker, we confirm the experimental observation that the wave group fissions into two packets.   The NFA simulations are able to predict a weak wave nonlinearity that occurs over one hundred wave periods.   The agreement between results of the numerical simulations and experimental measurements suggests that VOF simulations could be used to study the long-time evolution of a seaway complete with wave breaking.
 
\begin{figure*}
\begin{center}
\begin{tabular}{cccc}
$x/\lambda$ & Experiments & NFA & HOS \\
\raisebox{1.65cm}{4.88} & \includegraphics[width=0.15\linewidth]{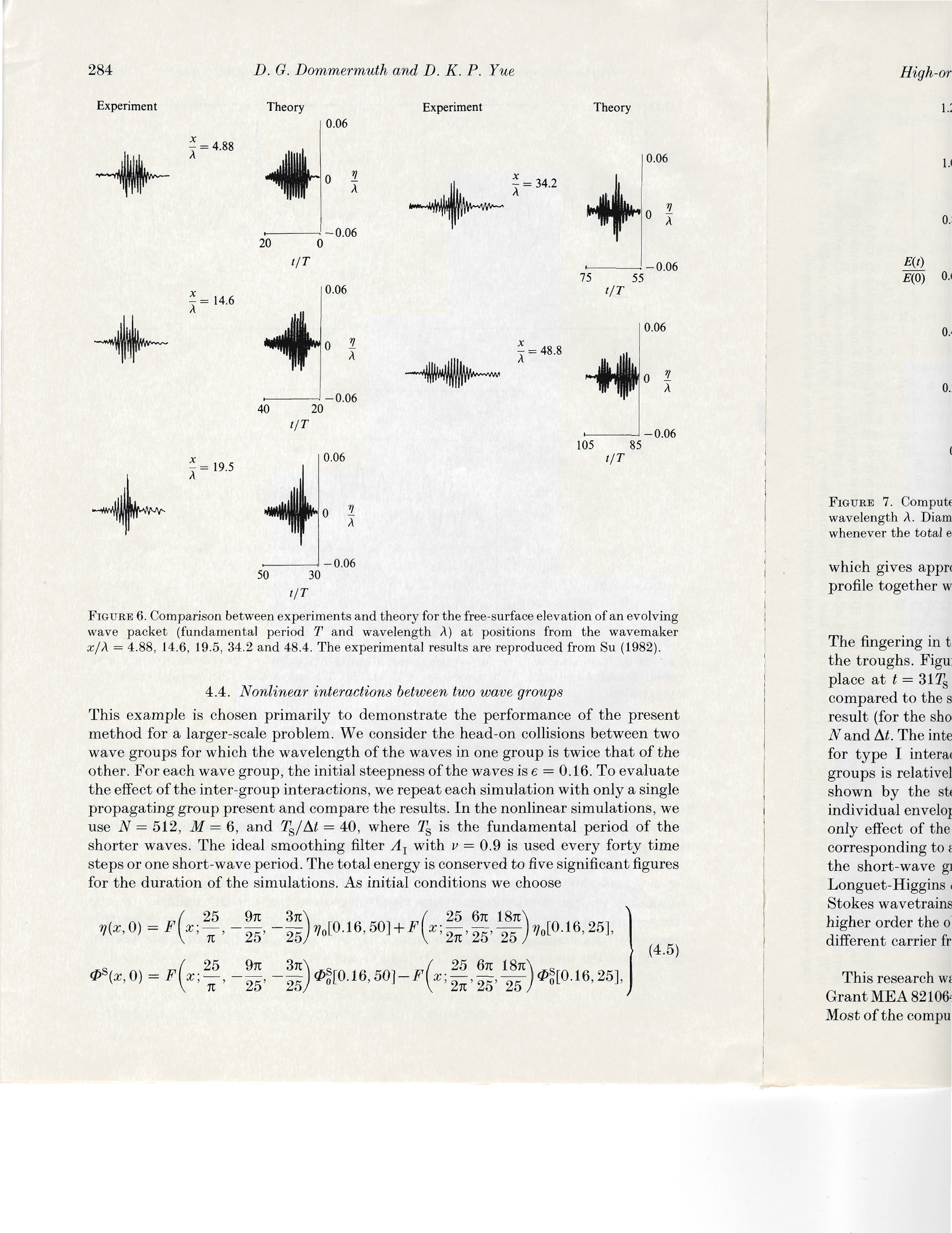}  & \raisebox{0.1cm}{\includegraphics[width=0.115\linewidth]{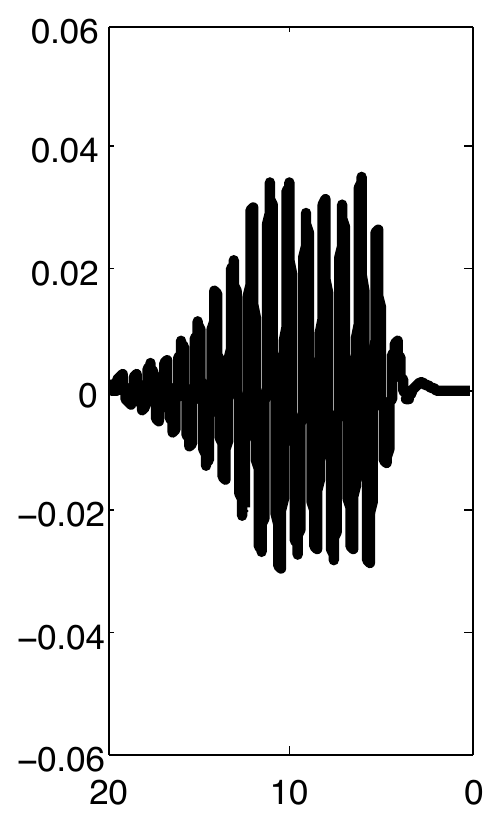}} & \includegraphics[width=0.15\linewidth]{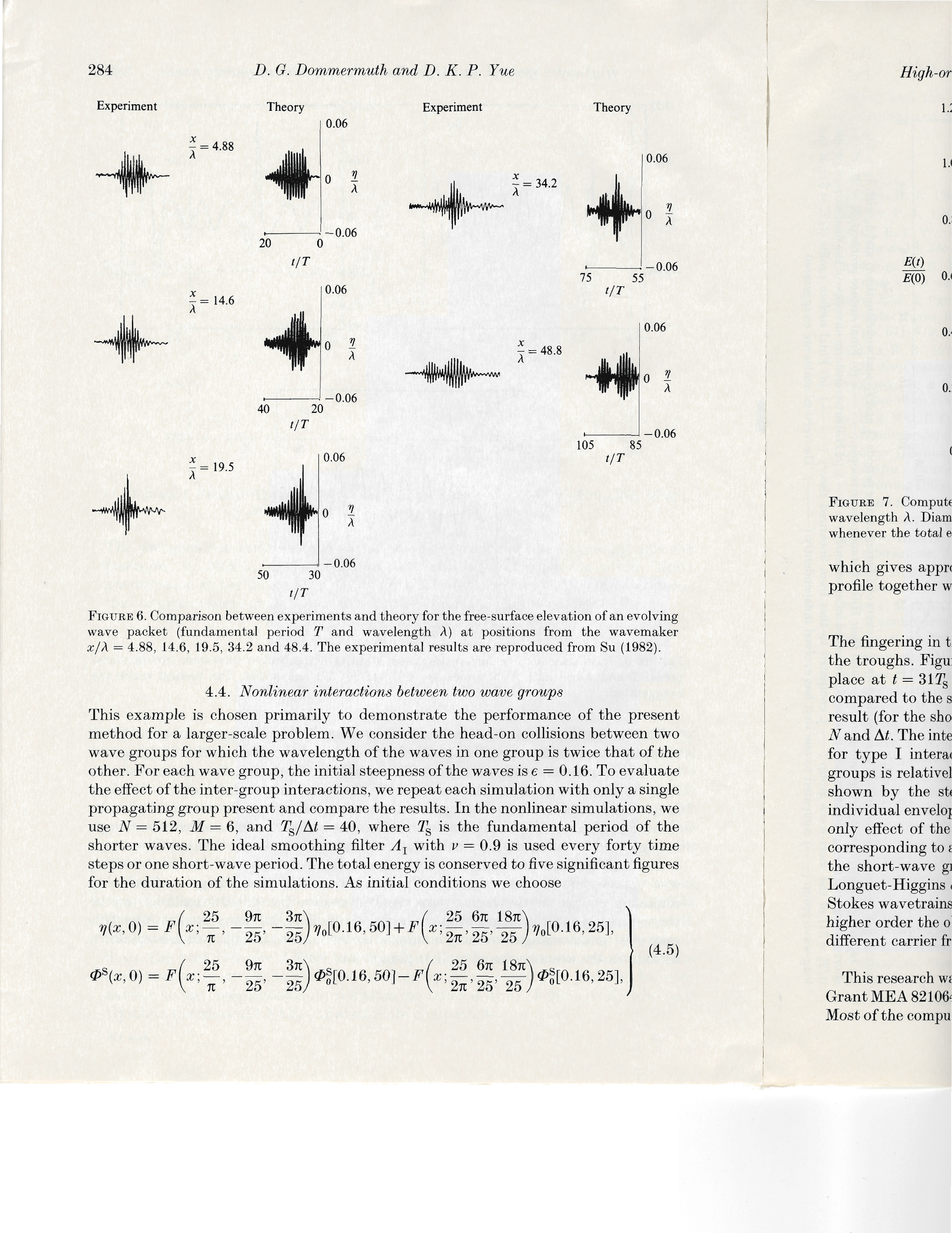}  \\ 
\raisebox{1.65cm}{14.6} & \includegraphics[width=0.15\linewidth]{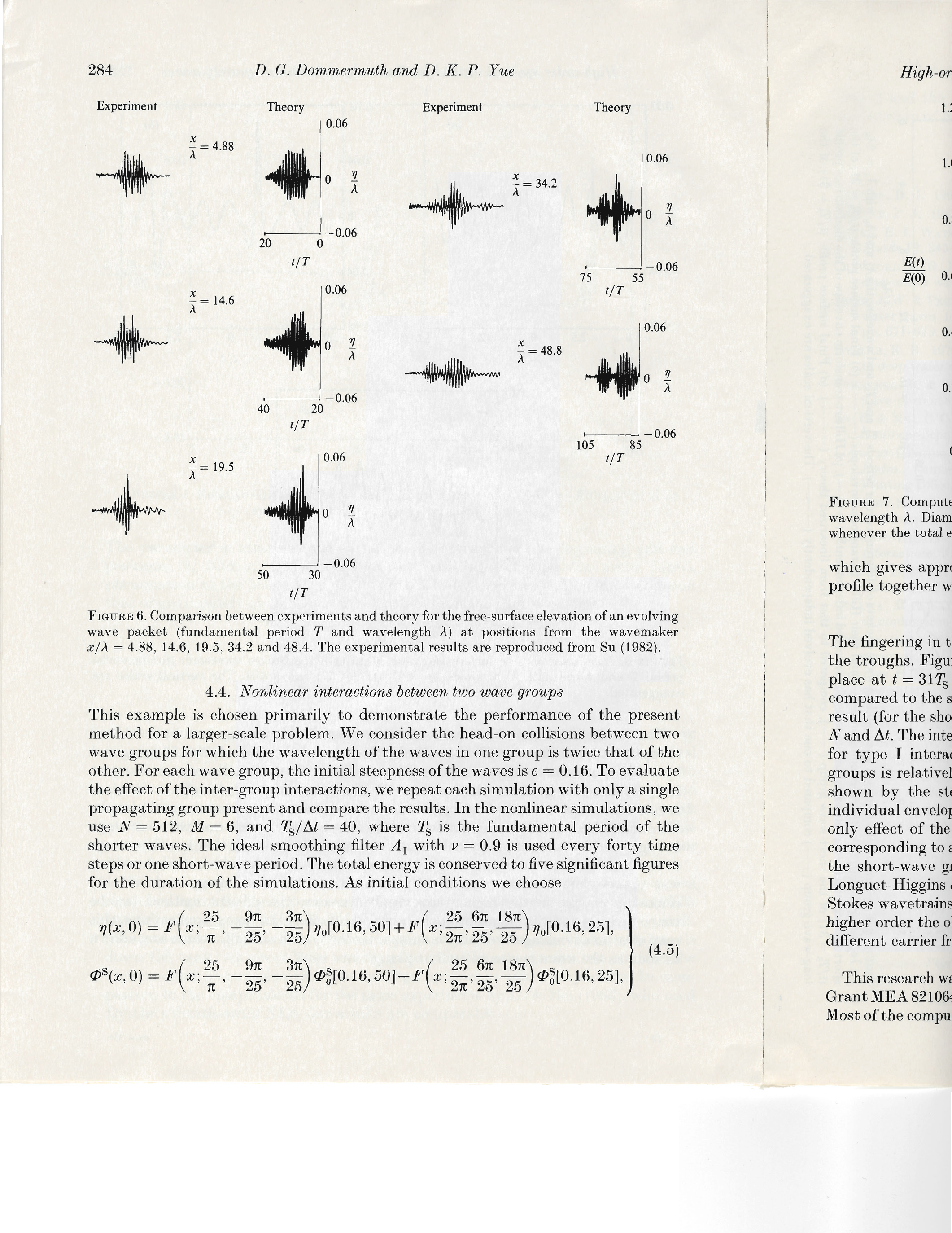}  & \raisebox{0.1cm}{\includegraphics[width=0.115\linewidth]{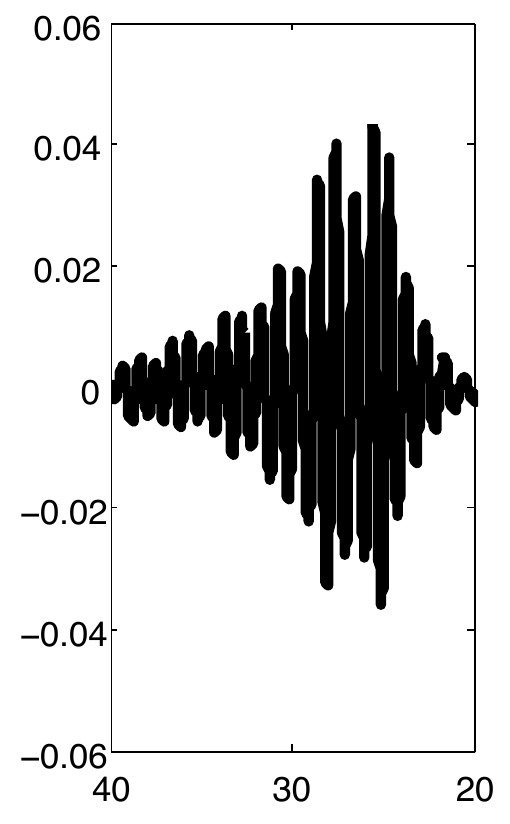}} & \includegraphics[width=0.15\linewidth]{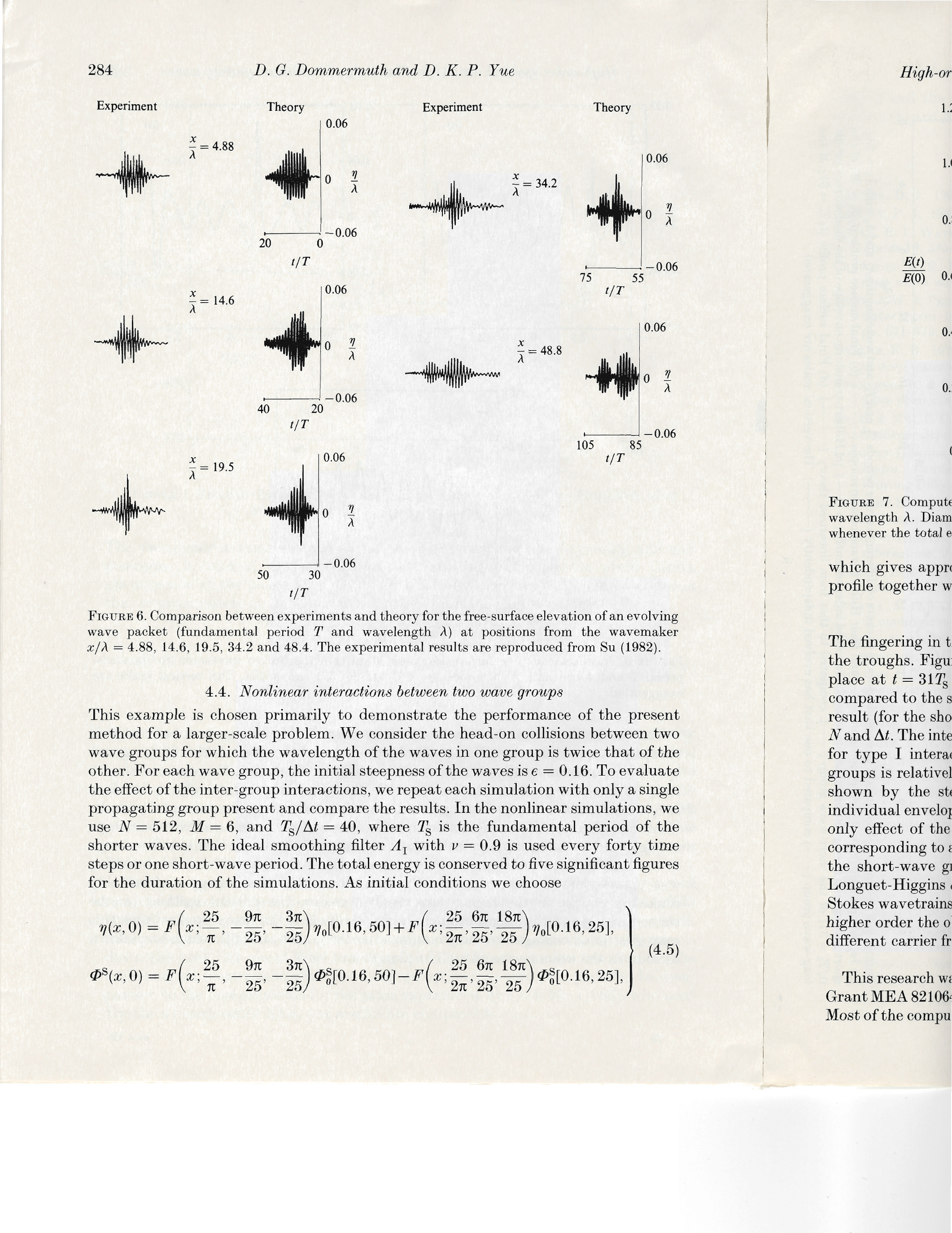}  \\ 
\raisebox{1.65cm}{19.5} & \includegraphics[width=0.15\linewidth]{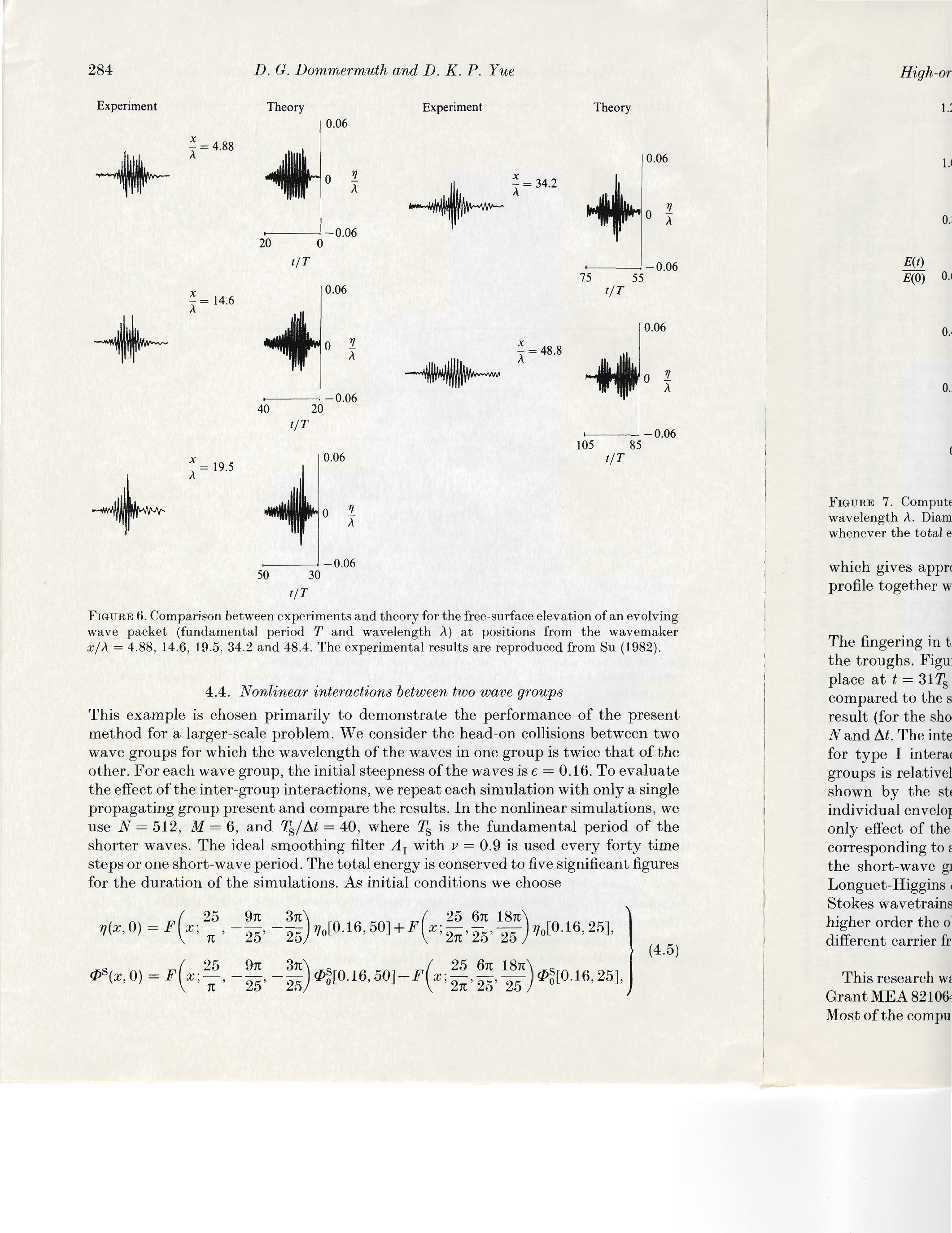}  & \raisebox{0.1cm}{\includegraphics[width=0.115\linewidth]{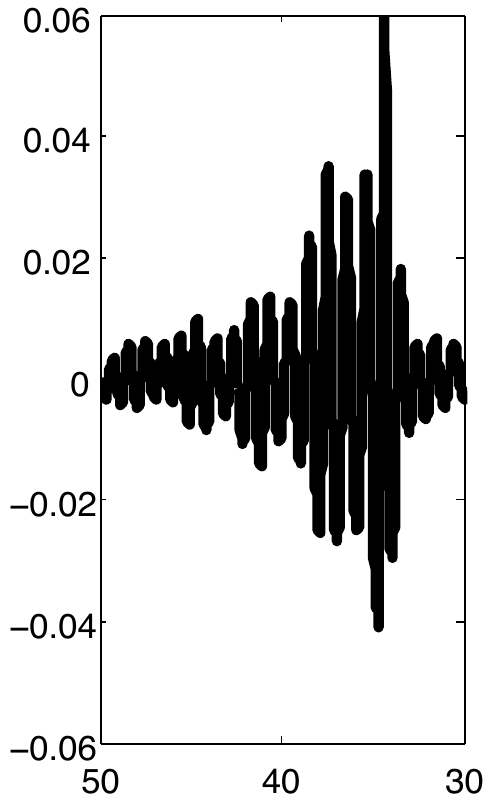}} & \includegraphics[width=0.15\linewidth]{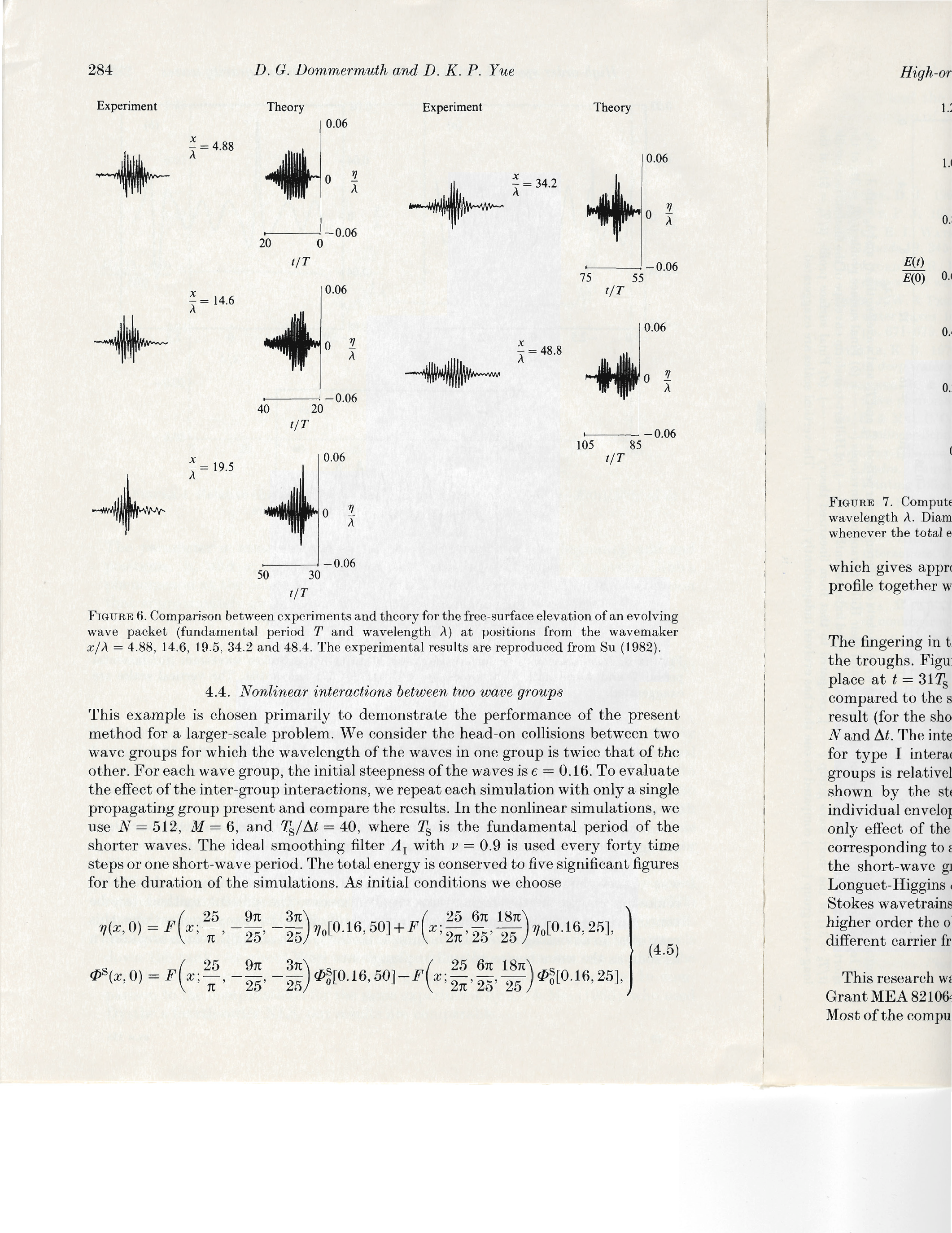}  \\ 
\raisebox{1.65cm}{34.2} & \includegraphics[width=0.15\linewidth]{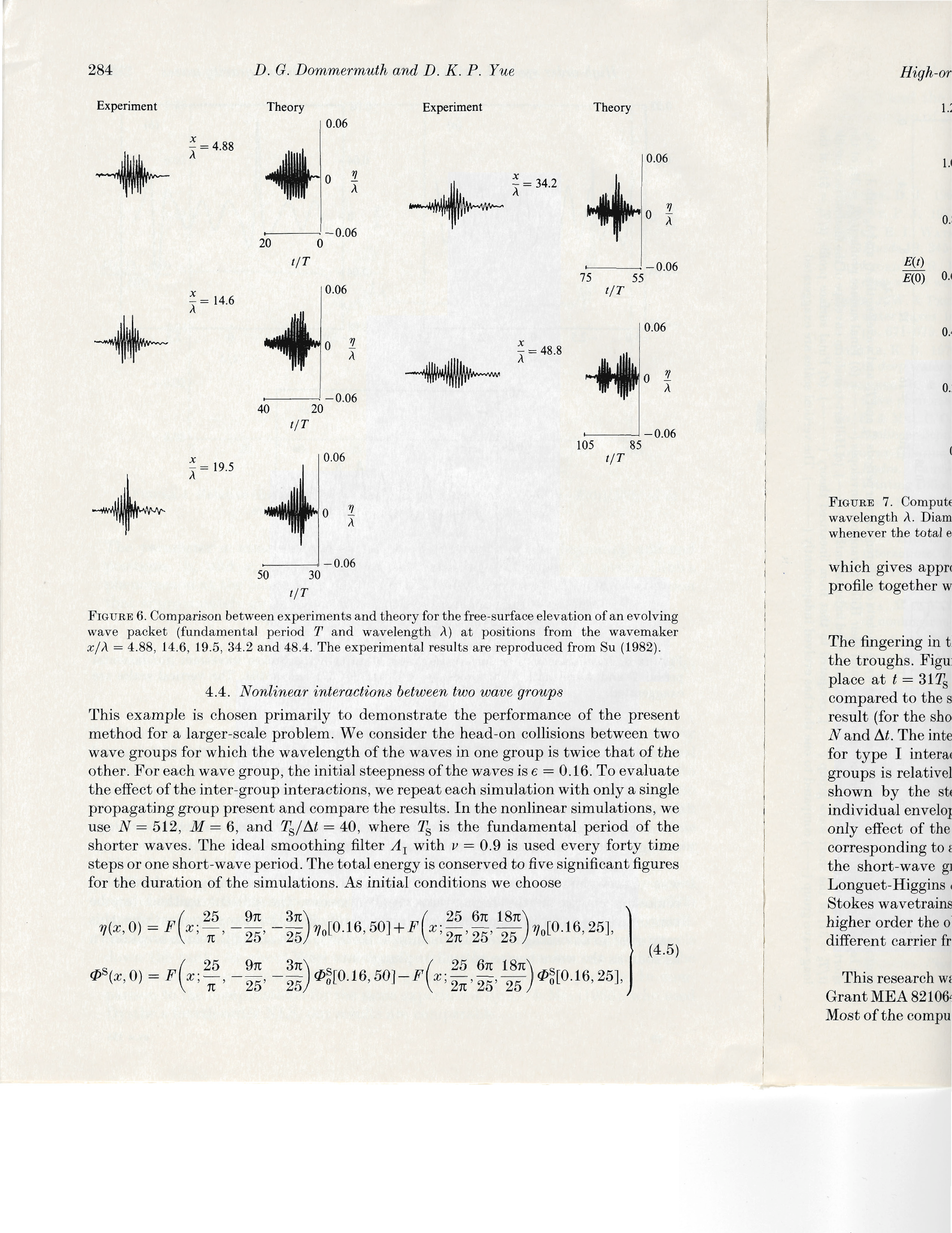}  & \raisebox{0.1cm}{\includegraphics[width=0.115\linewidth]{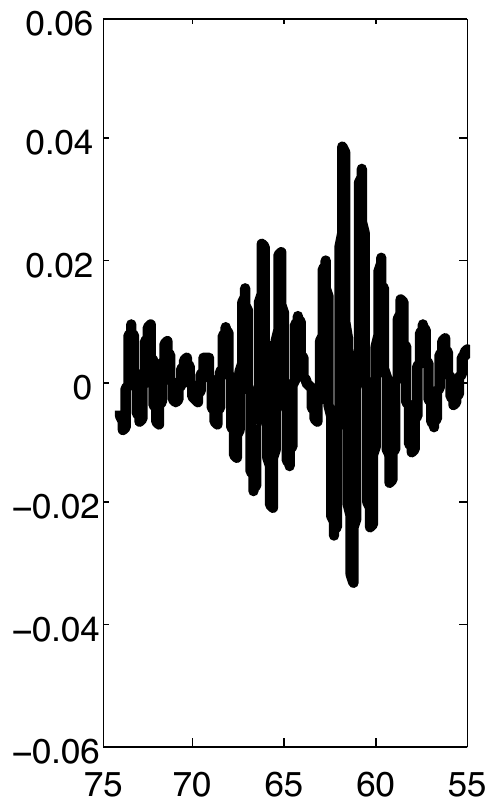}} & \includegraphics[width=0.15\linewidth]{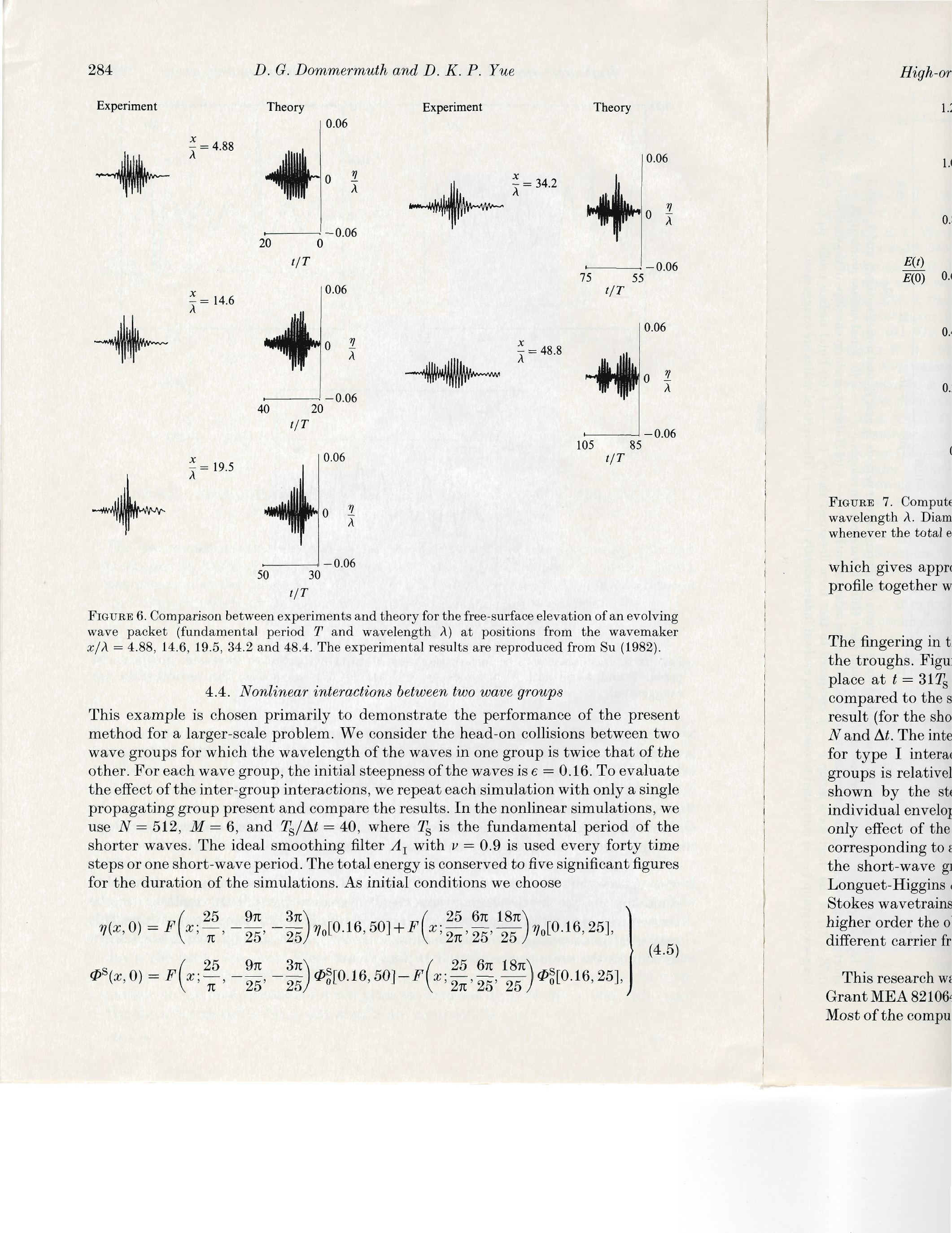}  \\
\raisebox{1.65cm}{48.8} & \includegraphics[width=0.15\linewidth]{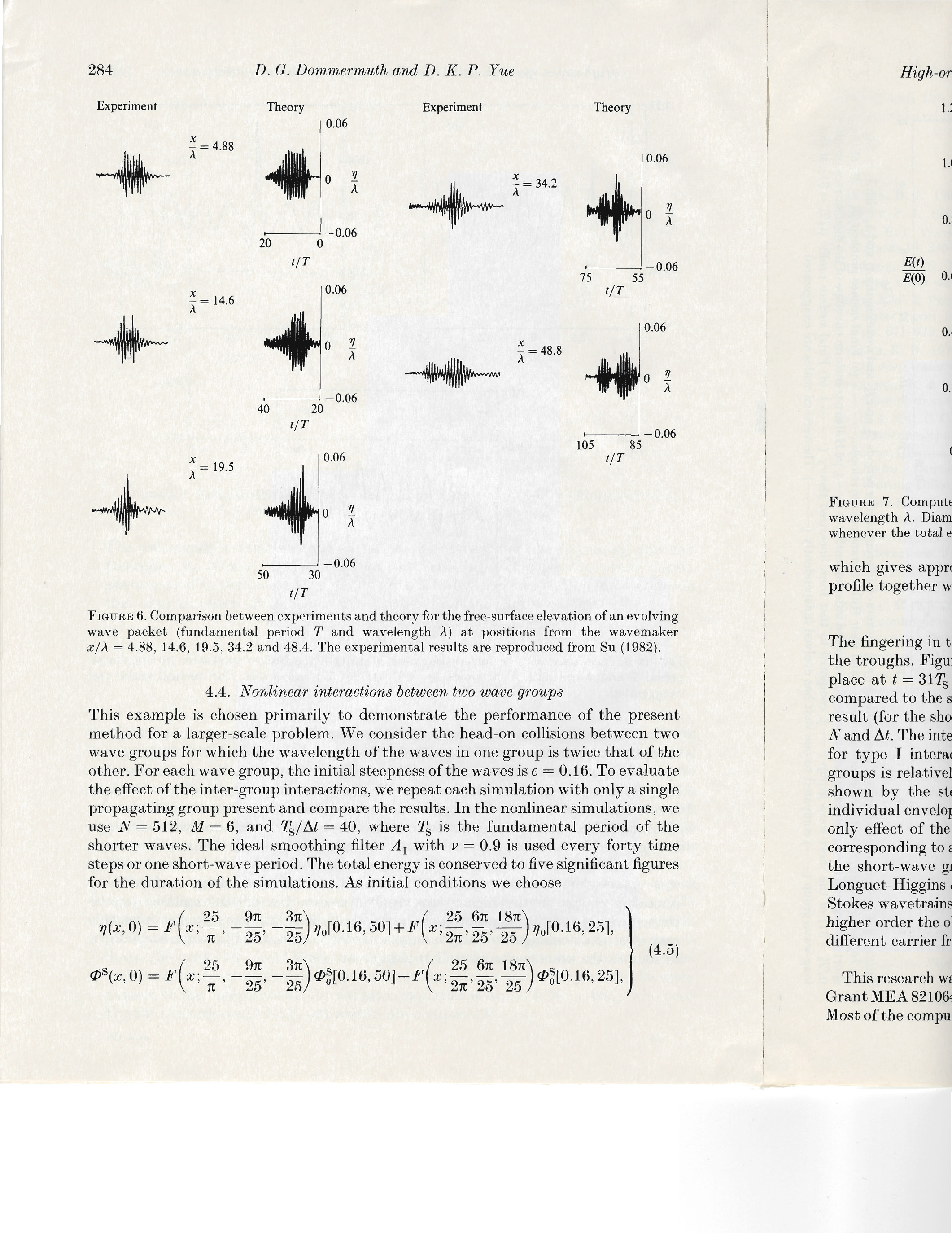}  &                                                                                                          & \includegraphics[width=0.15\linewidth]{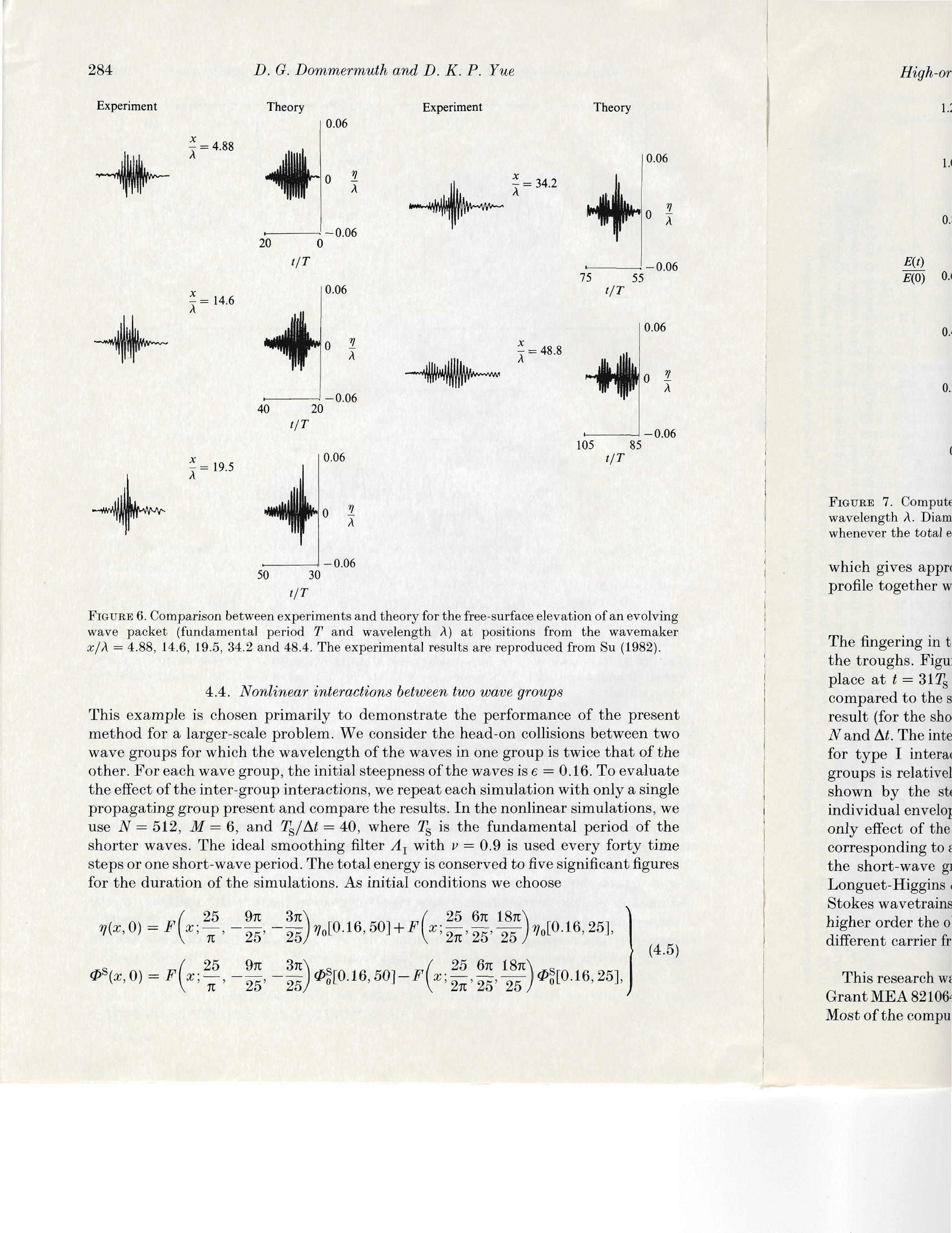}  \\
\end{tabular}
\end{center}
\caption{\label{fission} Fissioning of a wave packet.  $x/\lambda$ is the distance of the wave probe from the wavemaker.  $t/T$ is time normalized by the carrier wave period. }
\end{figure*}

\section{Conclusions}

The abilities of three different numerical methods for simulating nonlinear waves has been investigated.    HOS is very accurate and efficient, but HOS beaks down for simulations of broad-banded wave spectra.  FSM is capable of simulating broad-banded spectra, but it is computationally much more intensive than HOS.    TEMPEST and FREDYNE ship-motions codes could use interpolation of  pre-calculated FSM simulations as input to speed up computations.   NFA shows promise to simulate the wave breaking that occurs in high seastates.

\section{Acknowledgements}

The Office of Naval Research supports this research.  Dr. Patrick Purtell is the program manager.  This work is supported in part by a grant of computer time from the DOD High Performance Computing Modernization Program (http://www.hpcmo.hpc.mil/).  The numerical simulations have been performed on the Cray XT3 at the U.S. Army Engineering Research and Development Center (ERDC). 

\bibliography{28onr}
\bibliographystyle{28onr}

\end{document}